\documentclass[a4paper,11pt]{article}
\usepackage{a4wide,amsmath,amssymb,bbm}
\usepackage[english]{babel}

\renewcommand{\u}{\underline}

\makeatletter

\@addtoreset{equation}{section} \makeatother

\begin{document}

\hfill{Imperial-TP-LW-2015-01}

\vspace{40pt}

\begin{center}
{\LARGE{\bf On integrability of strings on symmetric spaces}}

\vspace{60pt}

Linus Wulff

\vspace{15pt}

{\it\small Blackett Laboratory, Imperial College, London SW7 2AZ, U.K.}\\

\vspace{120pt}

{\bf Abstract}

\end{center}
\noindent
In the absence of NSNS three-form flux the bosonic string on a symmetric space is described by a symmetric space coset sigma-model. Such models are known to be classically integrable. We show that the integrability extends also to cases with non-zero NSNS flux (respecting the isometries) provided that the flux satisfies a condition of the form $H_{abc}H^{cde}\sim R_{ab}{}^{de}$. We then turn our attention to the type II Green-Schwarz superstring on a symmetric space. We prove that if the space preserves some supersymmetry there exists a truncation of the full superspace to a supercoset space and derive the general form of the superisometry algebra. In the case of vanishing NSNS flux the corresponding supercoset sigma-model for the string is known to be integrable. We prove that the integrability extends to the full string by augmenting the supercoset Lax connection with terms involving the fermions which are not captured by the supercoset model. The construction is carried out to quadratic order in these fermions. This proves the integrability of strings on symmetric spaces supported by RR flux which preserve any non-zero amount of supersymmetry. Finally we also construct Lax connections for some supercoset models with non-zero NSNS flux describing strings in $AdS_{2,3}\times S^{2,3}\times S^{2,3}\times T^{2,3,4}$ backgrounds preserving eight supersymmetries.

\pagebreak 
\tableofcontents
\setcounter{page}{1}


\section{Introduction}
The discovery of integrability in the AdS/CFT holographic duality \cite{Maldacena:1997re} between string theory in $AdS_5\times S^5$ and $\mathcal N=4$ super Yang-Mills \cite{Minahan:2002ve,Bena:2003wd} has led to the exciting prospect of being able to solve completely both theories, at least in the large $N$ limit. In fact the spectral problem has been solved using the thermodynamic Bethe ansatz technique, see \cite{Beisert:2010jr} for a review. By now a number of other examples of $AdS$ backgrounds for which the string is integrable have been found and for some of them the dual CFT has also been identified. So far these examples have been constructed on a case by case basis. It would be very desirable to have a more systematic approach to finding integrable string backgrounds. Here we will attempt a small step in this direction by focusing on a simple, but important, class of backgrounds: symmetric spaces. In the context of supergravity we will mean by a symmetric space one where the geometry is that of a symmetric space and all supergravity fields respect the isometries. An incomplete\footnote{In the sense that the full moduli spaces for many $AdS_3$ and $AdS_2$ backgrounds were not determined.} classification of such symmetric space solutions to type IIB supergravity was given in \cite{FigueroaO'Farrill:2012rf} (a similar classification for eleven-dimensional supergravity was given in \cite{FigueroaO'Farrill:2011fj}). Since analyzing the integrability of this rather long list of backgrounds is still a difficult task we will make one further simplifying assumption, that the background preserve some non-zero amount of supersymmetry. A complete classification of supersymmetric symmetric space supergravity solutions does not appear to exist in the literature yet but it seems likely\footnote{For example in the eleven-dimensional case the list of backgrounds found in \cite{FigueroaO'Farrill:2011fj} was analyzed recently for supersymmetry in \cite{Hustler:2015lca} and no new examples were found.} that all examples with an $AdS$-factor are already known. All known backgrounds of this type arise from the near-horizon limit of (intersecting) brane configurations, either directly in ten dimensions or by reduction from eleven dimensions, see for example \cite{Boonstra:1998yu} for the $AdS\times S\times S\times T$ examples. These backgrounds are summarized in tables \ref{table:1} and \ref{table:2}. The first contains the backgrounds with only RR flux while the latter contains those with non-zero NSNS flux (in addition to RR flux in general). The string has been shown to be (classically) integrable for all of them.\footnote{For the cases without NSNS flux (table \ref{table:1}) integrability of the corresponding supercoset models follows from the construction of \cite{Bena:2003wd}. This was extended to the full string (to quadratic order in fermions) in \cite{Sorokin:2010wn,Sorokin:2011rr,Cagnazzo:2011at,Sundin:2012gc,Wulff:2014kja}. The cases with non-zero NSNS flux (table \ref{table:2}) were analyzed in \cite{Cagnazzo:2012se,Wulff:2014kja}.} 

\begin{table}[ht]
\begin{center}
\begin{tabular}{lc}
Background & Supersymmetries\\
\hline &\\
$AdS_5\times S^5$ & 32\\[10pt]
$AdS_4\times\mathbbm{CP}^3$ & 24\\[10pt]
$AdS_3\times S^3\times S^3\times S^1$ & 16\\[10pt]
$AdS_2\times S^2\times S^2\times T^4$ & 8\\[10pt]
\end{tabular}
\end{center}
\caption{Integrable supersymmetric symmetric space $AdS$-backgrounds: $H_{abc}=0$.}
\label{table:1}
\end{table}

\begin{table}[ht]
\begin{center}
\begin{tabular}{lc}
Background & Supersymmetries\\
\hline &\\
$AdS_3\times S^3\times S^3\times S^1$ & 16\\[10pt]
$AdS_3\times S^3\times S^2\times T^2$ & 8\\[10pt]
$AdS_3\times S^2\times S^2\times T^3$ & 8\\[10pt]
$AdS_2\times S^3\times S^3\times T^2$ & 8\\[10pt]
$AdS_2\times S^3\times S^2\times T^3$ & 8\\[10pt]
$AdS_2\times S^2\times S^2\times T^4$ & 8\\[10pt]
\end{tabular}
\end{center}
\caption{Integrable supersymmetric symmetric space $AdS$-backgrounds: $H_{abc}\neq0$.}
\label{table:2}
\end{table}

Note that all the $AdS\times S\times S\times T$ examples are actually one-parameter families of solutions. The parameter\footnote{Correspondingly the superisometry group involves the exceptional supergroup $D(2,1;\alpha)$.} $\alpha$ controls the relative radius of curvature of the different factors in the geometry. For a suitable choice of this parameter the radius of one of the spheres goes to infinity. In this way backgrounds of the form $AdS\times S\times T$ are obtained as special cases of the backgrounds listed in table \ref{table:1} and \ref{table:2}. The $AdS_3\times S^3\times S^3\times S^1$ and $AdS_2\times S^2\times S^2\times T^4$ solutions actually contain one more free parameter, $q$, which controls the amount of NSNS flux. This is why they appear in both tables. For $q=0$ they are supported by pure RR flux while for non-zero $q$ they are supported by a mix of NSNS and RR flux. The parameter $q$ comes from the possibility of performing an S-duality transformation on the corresponding pure RR backgrounds. Finally it is worth noting that the backgrounds in table \ref{table:2} with both two and three-dimensional factors in the geometry are related to the (pure RR) $AdS_3\times S^3\times S^3\times S^1$ background by one or more Hopf T-dualities \cite{Duff:1998cr} on the fiber of $S^3$ ($AdS_3$) viewed as a Hopf fibration over $S^2$ ($AdS_2$). The same is true for the $AdS_2\times S^2\times S^2\times T^4$ solution with maximal NSNS flux ($q=1$).

In this paper we will take a general approach and deal mostly with general symmetric space backgrounds. The only exception is when we discuss the integrability for the superstring with non-zero NSNS-flux for which we will have to specify to the backgrounds listed in table \ref{table:2}.

As a warm-up we will discuss the bosonic string on a symmetric space. We will find the general form of the isometry algebra and discuss the isometry conditions on the NSNS flux. We will also construct the Killing vectors and the conserved isometry Noether current of the string. In the absence of NSNS flux the bosonic string action is simply a symmetric space sigma-model which have been known for a long time to be integrable \cite{Eichenherr:1979ci}. We show that the integrability still holds in the presence of NSNS flux if it is of the form
\begin{equation}
H=\omega_3+\omega_2dy+dy^1dy^2dy^3\,,
\end{equation}
where $\omega_{2(3)}$ are invariant irreducible two(three)-forms and the $y$'s are coordinates of flat directions, and satisfies
\begin{equation}
H_{ab\u c}H^{\u{cde}}=\mathrm{const.}\times R_{ab}{}^{\u{de}}\,,\qquad H_{abc'}H^{c'\u{de}}=\mathrm{const.}\times R_{ab}{}^{\u {de}}\,,
\label{eq:HH-R}
\end{equation}
where we have split the index $a$ into curved directions $\u a$ and flat directions $a'$ and $R_{ab}{}^{cd}$ is the Riemann tensor. The Lax connection is essentially the same as appeared in \cite{Wulff:2014kja} in the context of the backgrounds in table \ref{table:2}. To my knowledge this observation has not appeared in the literature before.\footnote{Conditions for the integrability of a general non-linear sigma-model, including $B$-field, were written down in 
\cite{Mohammedi:2002xu,Mohammedi:2008vd} but they have only been solved in special cases.} These conditions on $H$ are likely related to supersymmetry. In fact they are satisfied for all the backgrounds in table \ref{table:2}. We should also note that there is one well-known non-supersymmetric symmetric space background which is known to be integrable namely $AdS_5\times\mathbbm{CP}^2\times S^1$, related by Hopf T-duality to $AdS_5\times S^5$ \cite{Duff:1998us}. For this background we have $H\sim\omega_2dy$ where $\omega_2$ is the K\"ahler form on $\mathbbm{CP}^2$, which does not satisfy the condition (\ref{eq:HH-R}). This means that this condition cannot be the most general one required for integrability.\footnote{It might be the most general condition if one requires the Lax connection to be constructed only from the isometries of the background (note that at least naively the Lax connection for the $AdS_5\times\mathbbm{CP}^2\times S^1$ string will involve the full 'hidden' $SO(4,2)\times SO(6)$-symmetry of $AdS_5\times S^5$ rather than the $SO(4,2)\times SU(3)\times U(1)$ isometries).}

Next we turn our attention to the (type II) superstring. We are interested in symmetric space backgrounds so we require there to be ten bosonic translational isometries and $n$ fermionic ones for a background that preserves $n$ supersymmetries. We show that by setting the $32-n$ fermions which are not associated to supersymmetries to zero the conditions on the background superfields to respect the isometries simplify considerably. In fact this $(10|n)$-dimensional sub-superspace is a supercoset space, as we prove by constructing the supergeometry. This proof also leads to the general form of the superisometry algebra for symmetric space backgrounds. We show that in the absence of NSNS flux, and for particular cases with NSNS flux, the $\mathbbm Z_2$-automorphism of the bosonic subalgebra extends to a $\mathbbm Z_4$-automorphism of the full super Lie algebra making the corresponding supercoset space a semisymmetric superspace \cite{Serganova:1983vp}. The semisymmetric supercosets relevant to string theory were classified in \cite{Zarembo:2010sg}. The only backgrounds of table \ref{table:2} which have this semisymmetric structure are the ones which also appear in table \ref{table:1}.

The semisymmetric supercoset sigma model describing (a truncation of) the string action in symmetric space backgrounds with zero NSNS flux are known to be integrable by the general construction of the Lax connection given in \cite{Bena:2003wd}. We show that the integrability extends to the full superstring by including also the $32-n$ fermions which do not correspond to supersymmetries and which are therefore absent in the supercoset formulation. This is done by extending the supercoset Lax connection by terms involving components of the superisometry Noether current and carried out to quadratic order in the non-coset fermions. The construction is essentially the same as was done for $AdS_4\times\mathbbm{CP}^3$ and $AdS_2\times S^2\times T^6$ in \cite{Cagnazzo:2011at}, but greatly simplified by working with components of the Noether current instead of the explicit expansion in non-coset fermions. Our construction requires that the background preserves some non-zero amount of supersymmetry. We comment on the problems with trying to generalize to non-supersymmetric backgrounds. Note that the extension of the integrability beyond the supercoset model to the full string is important even in cases where the supercoset model can be obtained by kappa symmetry gauge-fixing the full Green-Schwarz string. This is because the corresponding kappa symmetry gauge-fixing becomes inconsistent for many classical string configurations that one is typically interested in.

For backgrounds with non-zero NSNS flux there is no known general construction of a Lax connection. Instead we will focus on the backgrounds in table \ref{table:2} for which a Lax connection was found to quadratic order in all fermions in \cite{Wulff:2014kja}. Here we construct the Lax connection for the corresponding supercoset models (except for the $AdS_3\times S^3\times S^3\times S^1$ case which was covered in \cite{Cagnazzo:2012se}). We find that the remaining backgrounds in table \ref{table:2} can be described by a single Lax connection which is a rather straight-forward generalization of the Lax connection for the bosonic string with NSNS flux. Due to the construction of a Lax connection to quadratic order in all fermions in \cite{Wulff:2014kja} it is clear that it must be possible to extend this supercoset Lax connection by the non-coset fermions just as for the case of zero NSNS flux.

The outline of the paper is as follows. In section \ref{sec:bosonic} we describe the bosonic string on a symmetric space, the superisometry algebra, Killing vectors and the conditions on the NSNS flux. Section \ref{sec:bosonic-int} describes the Lax connection for the bosonic string in backgrounds that include NSNS flux satisfying eq. (\ref{eq:HH-R}). After this warm-up we turn to the full (type II) superstring in sec. \ref{sec:superstring}. We describe the conditions for the background to be a symmetric space and show that there exists a truncation to a supercoset subspace. The superisometry algebra, Killing vectors and Killing spinors are described and we comment on the special case of semisymmetric supercosets. We also describe the Noether current of the string associated to the superisometries. In section \ref{sec:integrability} we show how the standard Lax connection for the supercoset sigma model, in the case of zero NSNS flux, can be improved by terms involving the non-coset fermions to a Lax connection for the full superstring. We then construct Lax connections for the supercoset models corresponding to the backgrounds in table \ref{table:2} which have not previously been considered. We end with some conclusions. Appendix \ref{app:J} gives the expansion of the supergeometry and Noether current to quadratic order in the non-coset fermions and some relations satisfied by the components of the Noether current, which are needed in the proof of integrability, are derived.

\section{Bosonic string on a symmetric space}\label{sec:bosonic}
The bosonic string action is (we use 2d form notation and wedge products and pull-backs to the worldsheet $\Sigma$ are understood)
\begin{equation}
S=-T\int_\Sigma\left(\frac12*e^ae^b\eta_{ab}-B\right)\,,
\end{equation}
where $e^a$ are the vielbeins of the background and $B=\frac12e^be^aB_{ab}$ is the NSNS two-form potential with field strength $H=dB$. The equations of motion which follow from this action are
\begin{equation}
\nabla*e^a+\frac12e^ce^bH_{bc}{}^a=0\,.
\label{eq:bosonic-eom}
\end{equation}

We are interested in the case when the background geometry is that of a symmetric space. We will assume also that the $B$-field respects the symmetries of the background, which means that it can only transform by a gauge-transformation under the isometries. A symmetric space $\mathcal M$ can be described as a coset
\begin{equation}
\mathcal M=\frac{G}{H}\,,
\end{equation}
where $G$ has the interpretation as the isometry group of the space and $H$ is the isotropy group. Picking a coset representative $g\in G$ one defines the Maurer-Cartan one-form valued in the Lie algebra of $G$ as
\begin{equation}
J=g^{-1}dg\in\mathfrak{g}\,.
\label{eq:MC-form}
\end{equation}
By construction it satisfies the Maurer-Cartan equation\footnote{In our conventions the exterior derivative acts from the right, as is more convenient for the superspace discussion, and wedge-products will be left implicit.}
\begin{equation}
dJ-JJ=0\,.
\end{equation}
For a symmetric space the Lie algebra of the isometry group consists of generators of translations and rotations
\begin{equation}
\mathfrak{g}=\{M_{ab},\,P_a\}\,.
\end{equation}
Note that all the $P_a$'s must exist, i.e. the number of translational isometries is equal to the dimension of the space, while some of the $M_{ab}$'s may be absent depending on the space. We can then expand the Maurer-Cartan form in the generators as
\begin{equation}
J=\frac12\omega^{ab}M_{ab}+e^aP_a\,.
\label{eq:MCexp}
\end{equation}
The one-form coefficients $\omega^{ab}$ and $e^a$ are interpreted as the spin connection and vielbein of the background. A short calculation shows that the Maurer-Cartan equation $dJ-JJ=0$ becomes the statement that the torsion vanishes
\begin{equation}
T^a\equiv de^a+e^b\omega_b{}^a=0
\end{equation}
and the definition of the Riemann curvature tensor
\begin{equation}
R^{ab}\equiv d\omega^{ab}+\omega^{ac}\omega_c{}^b=\frac12e^de^cR_{cd}{}^{ab}\,,
\end{equation}
provided that the isometry algebra takes the familiar form
\begin{align}
&[M_{ab},P_c]=2\eta_{c[a}P_{b]}\,,\qquad
[P_a,P_b]=-\frac12R_{ab}{}^{cd}M_{cd}\,,
\nonumber\\
&[M_{ab},M_{cd}]=\eta_{ac}M_{bd}-\eta_{bc}M_{ad}-\eta_{ad}M_{bc}+\eta_{bd}M_{ac}\,.
\label{eq:bosonic-algebra}
\end{align}
Note that some care is required in interpreting the commutators involving $M_{ab}$. They are understood to be defined only when $M_{ab}$ exists, i.e. when the indices $a$ and $b$ refer to the same (irreducible) factor if the geometry is given by a direct product. We will also set to zero $M_{ab'}$ where primed indices denote flat directions, which has the consequence that the translation generators $P_{a'}$ of flat directions just generate decoupled $U(1)$'s. The non-zero components of $M_{ab}$ are precisely the same as the non-zero components of the spin connection $\omega^{ab}$, so the commutators hold for all components if all $M_{ab}$ are contracted with $\omega^{ab}$. We use this notation to avoid having to split the tangent space indices into ones for each irreducible component of the geometry.

Note also that the form of the algebra implies that there is a $\mathbbm Z_2$-automorphism
\begin{equation}
P_a\rightarrow-P_a\,,\qquad M_{ab}\rightarrow M_{ab}\,.
\end{equation}
This is usually taken as part of the definition of a symmetric space. 

Analyzing the Jacobi identities of the isometry algebra in eq. (\ref{eq:bosonic-algebra}) we find that they are equivalent to the following conditions on the Riemann tensor
\begin{equation}
R_{[abc]}{}^dP_d=0\,,\qquad \omega_{[a}{}^eR_{b]e}{}^{cd}=0\,.
\end{equation}
The first condition is just the familiar Bianchi identity for the curvature and the second (together with the fact that the curvature is constant) expresses the invariance of the curvature under the isometries.

Let us now consider the form of the isometry transformations of the coordinates. In the coset description isometry transformations are simply left-multiplication by a constant element of $G$, i.e. $g\rightarrow g_0g$ with $g_0\in G$. This clearly leaves the Maurer-Cartan form in eq. (\ref{eq:MC-form}) invariant. Consider an infinitesimal transformation with $g_0=e^{\epsilon^aP_a+\epsilon^{ab}M_{ab}}$ where $\epsilon^I=\{\epsilon^a,\,\epsilon^{ab}\}$ are constant infinitesimal parameters. This corresponds to a transformation of the coordinates
\begin{equation}
\delta x^m=\epsilon^IK_I{}^m=K_{(\epsilon)}^m\,,
\end{equation}
together with a local Lorentz transformation by $h\in H$. To see what the object $K_{(\epsilon)}^m$ is we use the fact that since the Maurer-Cartan form is invariant the vielbeins do not transform under this transformation, i.e.
\begin{equation}
0=\delta e^a=d\delta x^a-i_{\delta x}(e^b\omega_b{}^a)-e^bl_b{}^a=\nabla K_{(\epsilon)}^a-e^bi_{K_{(\epsilon)}}\omega_b{}^a-e^bl_b{}^a\,,
\label{eq:nablaKa}
\end{equation}
where we defined $K_{(\epsilon)}^a\equiv K_{(\epsilon)}^me_m{}^a$ and the anti-symmetric matrix $l_{ab}$ encoding the local Lorentz transformation. From this equation we find
\begin{equation}
\nabla^aK^b_{(\epsilon)}=i_{K_{(\epsilon)}}\omega^{ab}+l^{ab}\,,
\label{eq:lab}
\end{equation}
which implies in particular that
\begin{equation}
\nabla^{(a}K^{b)}_{(\epsilon)}=0\,,
\end{equation}
i.e. $K_{(\epsilon)}^a$ have the interpretation of Killing vectors. 

Consider next the transformation of the group element $g$. On the one hand we have
\begin{equation}
g^{-1}\delta g=g^{-1}(\epsilon^aP_a+\epsilon^{ab}M_{ab})g\,,
\end{equation}
and on the other hand we have, comparing to the definition of the Maurer-Cartan form in eqs. (\ref{eq:MC-form}) and (\ref{eq:MCexp}),
\begin{equation}
g^{-1}\delta g=\frac12\delta x^m\omega_m{}^{ab}M_{ab}+\delta x^me_m{}^aP_a=\frac12K^c_{(\epsilon)}\omega_c{}^{ab}M_{ab}+K^a_{(\epsilon)}P_a\,.
\end{equation}
Matching the $\epsilon^a$ and $\epsilon^{ab}$-terms in both equations we find
\begin{align}
\frac12K_a{}^b\omega_b{}^{cd}M_{cd}+K_a{}^bP_b=g^{-1}P_ag\,,
\qquad
\frac12K_{ab}{}^c\omega_c{}^{de}M_{de}+K_{ab}{}^cP_c=g^{-1}M_{ab}g\,.
\end{align}
Taking the trace of both equations with $P^d$, multiplying the last equation by $R_{fg}{}^{ab}$ and using $\mathrm{tr}(P_aP_b)=\eta_{ab}$, $\mathrm{tr}(M_{ab}P_c)=0$ and $R_{cd}{}^{ef}\mathrm{tr}(M_{ab}M_{ef})=-4\eta_{a[c}\eta_{d]b}$ (again the latter holds when all indices belong to the same irreducible component of the geometry) we find
\begin{align}
K_a{}^d=&\mathrm{tr}\big(g^{-1}P_agP^d\big)=[gP^dg^{-1}]_{P^a}
\\
R_{ab}{}^{ef}K_{ef}{}^d=&R_{ab}{}^{ef}\mathrm{tr}\big(g^{-1}M_{ef}gP^d\big)=-4[gP^dg^{-1}]_{M^{ab}}\,.
\end{align}
Using these equation we find
\begin{equation}
K^a\equiv P^bK_b{}^a-\frac14M^{cd}R_{cd}{}^{ef}K_{ef}{}^a=gP^ag^{-1}\,.
\end{equation}
This object $K^a$ is the Killing vector valued in the Lie algebra since it is obtained by replacing the parameters $\epsilon^I$ with the Lie algebra generators. Taking its exterior derivative and using the definition of the Maurer-Cartan form we find
\begin{equation}
dK_a=g[J,P_a]g^{-1}=g\big(\omega_a{}^bP_b+\frac12e^bR_{ab}{}^{cd}M_{cd}\big)g^{-1}\quad\Rightarrow\quad\nabla_aK_b=-\frac12R_{ab}{}^{cd}gM_{cd}g^{-1}\,.
\end{equation}
Comparing to eq. (\ref{eq:lab}) we read off the form of the (Lie algebra valued) local Lorentz transformation, $l^{ab}=g\left(-P^c\omega_c{}^{ab}-\frac12R_{ab}{}^{cd}M_{cd}\right)g^{-1}$. We will use these objects below to define the isometry Noether current of the string (valued in the isometry algebra).

The condition that the $B$-field respect the isometries is
\begin{equation}
d\lambda'_{(\epsilon)}=\delta B=\mathcal L_{\delta x}B=di_{K_{(\epsilon)}}B+i_{K_{(\epsilon)}}H\quad\Rightarrow\quad i_{K_{(\epsilon)}}H=d\lambda_{(\epsilon)}\,,
\end{equation}
where $\lambda'_{(\epsilon)}$ is the one-form gauge-parameter associated to gauge-transformations of $B$ and $\lambda_{(\epsilon)}=\lambda'_{(\epsilon)}-i_{K_{(\epsilon)}}B$. We can turn this into a Lie algebra valued equation by replacing $\{\epsilon^a,\,\epsilon^{ab}\}\rightarrow\{P^a,\,-\frac14M^{cd}R_{cd}{}^{ab}\}$ just as we did in the definition of the Lie algebra valued Killing vector $K^a$. It then reads
\begin{equation}
i_KH=d\lambda\,,\qquad K^a,\,\lambda\in\mathfrak{g}\,.
\end{equation}
In fact we can go further and prove that $H$ must be (covariantly) constant. We have
\begin{equation}
0=di_KH=
\frac12e^ce^bK^a\nabla H_{abc}
+\frac12e^ce^b\nabla K^aH_{abc}\,.
\end{equation}
Multiplying from the left by $g^{-1}$ and from the right by $g$ and using the form of $K^a$ and $\nabla K^a$ we find
\begin{equation}
\frac12e^de^ce^b\nabla_dH_{abc}P^a
-\frac14e^de^ce^bH_{abc}R_d{}^{aef}M_{ef}
=0\,.
\end{equation}
Finally taking the trace with $P_a$ and $M_{ab}$ and using the Bianchi identity we get\footnote{We have contracted the last equation with the spin connection to avoid having to split the indices in an awkward way. The content of the equation is unchanged.}
\begin{equation}
\nabla_aH_{bcd}=0\,,\qquad
\omega_{[b}{}^eH_{cd]e}=0\,.
\label{eq:bosonic-H-cond}
\end{equation}
Essentially the same as the conditions we found on the Riemann tensor. The first equation says that $H$ is covariantly constant and is the condition coming from invariance of $H$ with respect to the translational isometries. The second condition expresses the invariance of $H$ with respect to the rotational isometries. These conditions mean that $H$ must be built from invariant forms on the symmetric space. The only symmetric spaces with non-trivial invariant one-forms are pp-wave spaces \cite{FigueroaO'Farrill:2011fj} and since we are mainly interested in $AdS$-backgrounds we will not consider these. Then $H$ must be of the form
\begin{equation}
H=\omega_3+\omega_2dy+dy^1dy^2dy^3\,,
\label{eq:H-allowed}
\end{equation}
where $\omega_{2(3)}$ are invariant irreducible two(three)-forms and the $y$'s are coordinates of flat directions.
 
The (Lie algebra valued) isometry Noether current of the string can be constucted from the objects $K_a$ and $\lambda$. It is given by
\begin{equation}
\mathcal J=e^aK_a+*\lambda\,.
\end{equation}
Its conservation is easily verified
\begin{equation}
d*\mathcal J=\nabla*e^aK_a+*e^a\nabla K_a+d\lambda=\nabla*e^aK_a+i_KH=0\,,
\end{equation}
where we have used the equations of motion eq. (\ref{eq:bosonic-eom}), the Killing vector equation $\nabla_{(a}K_{b)}=0$ and the fact that $d\lambda=i_KH$. We can find a more explicit form of the one-form $\lambda$ as follows. Multiplying the equation $d\lambda=i_KH$ with $g^{-1}$ and $g$ from the left and right we get
\begin{equation}
i_PH=d[g^{-1}\lambda g]-[g^{-1}\lambda g,J]\,.
\end{equation}
A solution to this equation is easily seen, using eq. (\ref{eq:bosonic-H-cond}), to be given by
\begin{equation}
\lambda=\frac14(e^aH_{abc}+e^{a'}H_{a'bc})gM^{bc}g^{-1}+\lambda^{a'}P_{a'}
\,,\qquad
d\lambda^{a'}=\frac12e^ce^bH^{a'}{}_{bc}\,.
%
\end{equation}

\section{Integrability for the bosonic string}\label{sec:bosonic-int}
We want to construct a one-parameter family of flat connections such that the flatness condition is equivalent to the equations of motion for the string eq. (\ref{eq:bosonic-eom}). Let us start with the case of zero NSNS flux. It this case the Lax connection is well know and takes the form
\begin{equation}
L_0=\frac12\omega^{ab}M_{ab}+(1+\alpha)e^aP_a+\beta*e^aP_a\,,
\end{equation}
where the two parameters $\alpha$ and $\beta$ will be related by the requirement of flatness. Computing the curvature of this connection we get
\begin{align}
dL_0-L_0L_0=&
\frac12R^{ab}M_{ab}
+\beta\nabla*e^aP_a
-\frac12\big((1+\alpha)^2-\beta^2\big)e^ae^b[P_a,P_b]
\nonumber\\
=&
\beta\nabla*e^aP_a
-\frac12\big(\alpha^2+2\alpha-\beta^2\big)R^{ab}M_{ab}\,.
\end{align}
Where we used the form of the isometry algebra in eq. (\ref{eq:bosonic-algebra}). We see that if we take the parameters $\alpha$ and $\beta$ to be related by the equation
\begin{equation}
\beta^2=\alpha^2+2\alpha\,,
\label{eq:alpha-beta-rel}
\end{equation}
the curvature of $L_0$ vanishes precisely on the equations of motion. We have therefore constructed a one-parameter family of flat connections whose flatness is equivalent to the equations of motion, i.e. a Lax connection. Infinitely many conserved charges can now be constructed from the monodromy of this Lax connection.

Next we move on to the case of non-zero NSNS flux. We will first consider separately the different possible forms of $H$ in eq. (\ref{eq:H-allowed}) before going to the general case.

\subsection{$H=\omega_3$}
We take $H$ to be an (irreducible) invariant three-form of the symmetric space in question. There are now two terms\footnote{We could also consider terms of higher order in $H$ but we will not do so here.} we can add to the Lax connection (the underlined indices denote curved directions to emphasize that $H_{abc}$ has no legs along flat directions)
\begin{equation}
L_{\omega_3}=\frac12\omega^{ab}M_{ab}+(1+\alpha)e^aP_a+\beta*e^aP_a+\frac{\gamma}{4}e^{\u a}H_{\u{abc}}M^{\u{bc}}+\frac{\delta}{4}*e^{\u a}H_{\u{abc}}M^{\u{bc}}\,,
\end{equation}
with $\gamma$ and $\delta$ to be determined. Calculating the curvature of this connection we find
\begin{align}
dL_{\omega_3}-L_{\omega_3}L_{\omega_3}
=&
\beta\big(\nabla*e^a+\frac12e^ce^bH_{bc}{}^a\big)P_a
+\frac{\delta}{4}\big(\nabla*e^{\u a}+\frac12e^{\u e}e^{\u d}H_{\u{de}}{}^{\u a}\big)H_{\u{abc}}M^{\u{bc}}
\nonumber\\
&{}
+\frac12\big((1+\alpha)\gamma-(1+\delta)\beta\big)e^{\u b}e^{\u a}H_{\u{abc}}P^{\u c}
-\frac12\big(\alpha^2+2\alpha-\beta^2\big)R^{ab}M_{ab}
\nonumber\\
&{}
+\frac18(\delta^2-\gamma^2)e^{\u e}e^{\u d}H^{\u a}{}_{\u{db}}H_{\u{cea}}M^{\u{bc}}
-\frac{\delta}{8}e^{\u e}e^{\u d}H_{\u{ade}}H^{\u{abc}}M_{\u{bc}}\,,
\end{align}
where the first two terms are proportional to the equations of motion. The last term has a different structure than the others which means that either $\delta=0$ or we impose a condition on $H$ so that the last term becomes related to one of the previous two terms. Note that if we take $\delta=0$ we still need some condition on $H$ since $\gamma$ cannot also vanish. The conditions we need in the two cases turn out to be essentially equivalent. We will therefore allow $\delta\neq0$ and impose the following condition on $H$
\begin{equation}
H^{\u a}{}_{\u{d[e}}H_{\u{bc]a}}=0\,.
%
\label{eq:C3-H-cond}
\end{equation}
We will discuss this condition further at the end of the section. The last two terms in the expression for the curvature now combine to one and we are left with
\begin{align}
dL_{\omega_3}-L_{\omega_3}L_{\omega_3}
=&
\beta\big(\nabla*e^a+\frac12e^ce^bH_{bc}{}^a\big)P_a
+\frac{\delta}{4}\big(\nabla*e^{\u a}+\frac12e^{\u e}e^{\u d}H_{\u{de}}{}^{\u a}\big)H_{\u{abc}}M^{\u{bc}}
\nonumber\\
&{}
+\frac12\big((1+\alpha)\gamma-(1+\delta)\beta\big)e^{\u b}e^{\u a}H_{\u{abc}}P^{\u c}
-\frac12\big(\alpha^2+2\alpha-\beta^2\big)R^{ab}M_{ab}
\nonumber\\
&{}
+\frac18(\delta^2+2\delta-\gamma^2)e^{\u e}e^{\u d}H^{\u a}{}_{\u{db}}H_{\u{cea}}M^{\u{bc}}\,.
\end{align}
The last three terms now vanish provided that we take
\begin{equation}
\gamma=\beta\,,\quad\delta=\alpha\,,
\end{equation}
with $\alpha$ and $\beta$ subject to the relation we found before, eq. (\ref{eq:alpha-beta-rel}). Again the two remaining terms vanish precisely on the equations of motion completing the proof of integrability for this case.

\subsection{$H=\omega_2dy$}
In this case we will find that we need to split the curved indices $\u a$ into $(\hat a,\,\tilde a)$ where the hatted indices denote directions with $H$-flux, whose non-zero components are of the form $H_{\hat a\hat bc'}$. We will simplify the discussion here by only considering the terms in the Lax connection with hatted indices. The full Lax connection will be considered in the next section. Again there are two $H$-dependent terms that are natural to add to the Lax connection and we take
\begin{equation}
L_{\omega_2}=
\frac12\omega^{\hat a\hat b}M_{\hat a\hat b}+(1+\hat\alpha)e^{\hat a}P_{\hat a}+\hat\beta*e^{\hat a}P_{\hat a}
+\frac{\hat\gamma}{4}e^{a'}H_{a'\hat b\hat c}M^{\hat b\hat c}
+\frac{\hat\delta}{4}*e^{a'}H_{a'\hat b\hat c}M^{\hat b\hat c}\,,
\end{equation}
with $\hat\alpha,\,\hat\beta,\,\hat\gamma,\,\hat\delta$ to be determined. We have anticipated the fact that the coefficients $\alpha,\,\beta$ will now be different from before by denoting them with a hat. The curvature becomes
\begin{align}
dL_{\omega_2}-L_{\omega_2}L_{\omega_2}=&
\hat\beta\big(\nabla*e^{\hat a}+\frac12e^ce^bH_{bc}{}^{\hat a}\big)P_{\hat a}
+\frac{\hat\delta}{4}\big(\nabla*e^{a'}+\frac12e^ee^dH_{de}{}^{a'}\big)H_{a'\hat b\hat c}M^{\hat b\hat c}
\nonumber\\
&{}
+\frac12[(1+\hat\alpha)\hat\gamma-\hat\beta\hat\delta-2\hat\beta]e^{\hat b}e^{a'}H_{a'\hat b\hat c}P^{\hat c}
-\frac12[(1+\hat\alpha)\hat\delta-\hat\beta\hat\gamma]*e^{\hat b}e^{a'}H_{a'\hat b\hat c}P^{\hat c}
\nonumber\\
&{}
-\frac12\big(\hat\alpha^2+2\hat\alpha-\hat\beta^2\big)R^{\hat a\hat b}M_{\hat a\hat b}
-\frac{\hat\delta}{8}e^{\hat e}e^{\hat d}H_{a'\hat d\hat e}H^{a'\hat b\hat c}M_{\hat b\hat c}
\nonumber\\
&{}
+\frac18(\hat\gamma^2-\hat\delta^2)e^{a'}e^{d'}H_{a'\hat b\hat c}H^{\hat b}{}_{d'\hat e}M^{\hat c\hat e}\,.
\end{align}
The last term involves $H_{d'\hat e[\hat b}M_{\hat c]}{}^{\hat e}$ which vanishes due to the isometry condition in eq. (\ref{eq:bosonic-H-cond}). The first two terms are proportional to the equations of motion, however the third and fourth terms must clearly vanish. This implies that we must have $\hat\delta\neq0$ and the only way to have a non-trivial solution is then if the fifth and sixth term combine to one. This requires the following condition on $H$
\begin{equation}
H_{a'\hat d\hat e}H^{a'\hat b\hat c}=cR_{\hat d\hat e}{}^{\hat b\hat c}\,,
\label{eq:C2-H-cond}
\end{equation}
for some constant $c$. Requiring this we find the following solution\footnote{This solution does not reduce to the standard one when $H\rightarrow0$. A solution with this property can be found by taking $\alpha\rightarrow2\beta^2$ and $\beta\rightarrow2(1+\alpha)\beta$ which gives
$$
1+\hat\alpha=\sqrt{1-c\beta^2}\,(1+\alpha)\,,\quad
\hat\beta=\sqrt{1-c\beta^2}\,\beta\,,\quad
\hat\gamma=2(1+\alpha)\beta\,,\quad
\hat\delta=2\beta^2\,.
$$
\label{foot:alt-hatted}
}
\begin{equation}
\hat\gamma=\beta\,,\quad\hat\delta=\alpha\,,\quad\hat\beta=\eta\beta\,,\quad1+\hat\alpha=\eta(2+\alpha)\,,\qquad\eta=\frac12\sqrt{\frac{2-c\alpha}{2+\alpha}}\,,
\label{eq:alphahat}
\end{equation}
where $\alpha$ and $\beta$ satisfy the same equation as before, eq. (\ref{eq:alpha-beta-rel}). Note that in the special case $c=-1$ this simplifies as then $\eta=\frac12$. Again we see that flatness of the connection is equivalent to the equations of motion.

\subsection{General case}
Let us combine the cases considered so far into one general Lax connection for backgrounds with $H$ of the form in eq. (\ref{eq:H-allowed})
\begin{equation}
L=
\frac12\omega^{ab}M_{ab}+(1+\alpha_{[a]})e^aP_a+\beta_{[a]}*e^aP_a
+\frac{\beta}{4}e^aH_{abc}M^{bc}
+\frac{\alpha}{4}*e^aH_{abc}M^{bc}
+\beta\lambda^{a'}P_{a'}\,.
\label{eq:Lax}
\end{equation}
Note that we have added the last term which is needed to make the $P_{a'}$-terms in the curvature proportional to the equations of motion in the case when $H$ has a leg along the flat directions, $H=\omega_2dy$ (recall that $d\lambda^{a'}=\frac12e^ce^bH_{bc}{}^{a'}$ and that primed indices denote flat directions). To shorten the expression we have introduced a notation with $\alpha_{[a]},\,\beta_{[a]}$ depending on the index $a$ defined as
\begin{equation}
\alpha_{[a]}=
\left\{
\begin{array}{cc}
\hat\alpha & (a=\hat a)\\
\alpha & (a=\tilde a,\,a')
\end{array}
\right.\,,\qquad
\beta_{[a]}=
\left\{
\begin{array}{cc}
\hat\beta & (a=\hat a)\\
\beta & (a=\tilde a,\,a')
\end{array}
\right.\,,
\end{equation}
e.g.
\begin{equation}
\beta_{[a]}*e^aP_a=\hat\beta*e^{\hat a}P_{\hat a}+\beta*e^{\tilde a}P_{\tilde a}+\beta*e^{a'}P_{a'}\,.
\end{equation}
Recall that the hatted indices denote directions with non-zero NSNS flux of the form $H_{\hat a\hat bc'}$. Note that $M_{\hat a\tilde b}=0$ (and similarly for the spin connection and curvature) by the condition that $H$ respect the isometries in eq. (\ref{eq:bosonic-H-cond}) which implies $H_{a'\hat b\hat c}\omega^{\hat c\tilde d}=0$.  Using this fact together with the conditions on $H$ in eqs. (\ref{eq:C3-H-cond}) and (\ref{eq:C2-H-cond}) the curvature of this connection becomes, by a very similar calculation to the ones already performed,
\begin{align}
dL-LL=&
\beta_{[a]}(\nabla*e^a+\frac12e^ce^bH_{bc}{}^a)P_a
+\frac{\alpha}{4}(\nabla*e^a+\frac12e^ee^dH_{de}{}^a)H_{abc}M^{bc}
-\frac{\alpha}{8}e^{e'}e^{d'}H_{d'e'a'}H^{a'\hat b\hat c}M_{\hat b\hat c}\,.
\end{align}
The last term must vanish which implies a further condition on $H$
\begin{equation}
H_{d'e'a'}H^{a'\hat b\hat c}=0\,.
\label{eq:R3-H-cond}
\end{equation}
The remaining terms vanish precisely on the equations of motion demonstrating the integrability.

We can summarize the conditions on $H$ for the flatness of the connection in eq. (\ref{eq:Lax}) to be equivalent to the equations of motion as follows
\begin{equation}
H_{ab\u c}H^{\u{cde}}=c_1R_{ab}{}^{\u{de}}\,,\qquad H_{abc'}H^{c'\u{de}}=c_2R_{ab}{}^{\u {de}}\,,
\label{eq:HH-R2}
\end{equation}
where $c_{1,2}$ are constants. Note that the condition in eq. (\ref{eq:C3-H-cond}) follows from the Bianchi identity for the Riemann tensor. As mentioned in the introduction these conditions are probably related to supersymmetry. Indeed there is one well known integrable example which does not satisfy these conditions namely strings in $AdS_5\times\mathbbm{CP}^2\times S^1$, but this example is also non-supersymmetric (see the discussion in the introduction). It would be interesting if a more general construction could be found but it is not clear if this is possible if the Lax connection is built only from the isometries of the background.

Rather than dwell on these issues we will turn our attention to the topic of main interest here, namely the integrability of the superstring on a symmetric space.

\section{Type II superstring on a symmetric space}\label{sec:superstring}
We will consider the type II Green-Schwarz superstring. The action is
\begin{equation}
S=-T\int_\Sigma\left(\frac12*E^aE^b\eta_{ab}-B\right)\,,
\end{equation}
where $E^a$ are the supervielbeins of the background and $B=\frac12E^bE^aB_{ab}$ is the NSNS two-form potential superfield (pulled back to the worldsheet $\Sigma$). These fields depend on the ten bosonic coordinates $x^m$ and the 32 fermionic coordinates $\Theta$ of the superspace background (the explicit form of the action is known up to fourth order in  $\Theta$ for a general background \cite{Wulff:2013kga}). The equations of motion which follow from this action are
\begin{equation}
\nabla*E^a+\frac12E^CE^BH_{BC}{}^a=0\,,\qquad*E^aE^BT_{B\alpha}{}^c\eta_{ac}+\frac12E^CE^BH_{BC\alpha}=0\,.
\end{equation}
Using the superspace constraints on the torsion and $H$
\begin{equation}
T^a=-\frac{i}{2}E\Gamma^aE\,,\qquad H=-\frac{i}{2}E^a\,E\Gamma_a\Gamma_{11}E+\frac{1}{3!}E^cE^bE^aH_{abc}\,,
\end{equation}
these reduce to
\begin{equation}
\nabla*E^a-\frac{i}{2}E\Gamma^a\Gamma_{11}E+\frac12E^cE^bH_{bc}{}^a=0\,,\qquad*E^a\,(\Gamma_aE)-E^a\,(\Gamma_a\Gamma_{11}E)=0\,.
\label{eq:eom}
\end{equation}
Here, and in the following, we write the expressions that apply to type IIA. The corresponding expressions for type IIB are obtained by replacing the 32-component fermion with a doublet of 16-component ones and replacing the 32-component gamma-matrices by 16-component Weyl blocks
\begin{equation}
E\rightarrow E^i\,,\qquad\Gamma^a\rightarrow\gamma^a\,,\qquad\Gamma_{11}\rightarrow\sigma^3_{ij}\,.
\label{eq:IIAtoIIB}
\end{equation}
For more details we refer the reader to the appendix of \cite{Wulff:2013kga} whose conventions we use.

The superstring action is invariant under the isometry transformations with parameters $\{\epsilon^a,\,\epsilon^{ab}\}$, just as in the bosonic case. It is also invariant under any supersymmetries preserved by the background, whose parameters we denote $\epsilon^{\hat\alpha}$. These combine into the superisometry transformations ($z^M=(x^m,\,\Theta^\mu)$)
\begin{equation}
\delta z^M=\epsilon^aK_a{}^M+\epsilon^{ab}K_{ab}{}^M+\epsilon^{\hat\alpha}K_{\hat\alpha}{}^M=K_{(\epsilon)}^M\,.
\end{equation}
Since we want to describe a (ten-dimensional) symmetric space preserving $n$ supersymmetries we have by definition 
\begin{equation}
\mathrm{rank}\,\,K_a{}^m=10\qquad\mbox{and}\qquad\mathrm{rank}\,\,K_{\hat\alpha}{}^\mu=n\,.
\end{equation}
Defining $K_{(\epsilon)}^A=K_{(\epsilon)}^ME_M{}^A$ this means that $K_a{}^b$ and $K_{\hat\alpha}{}^{\hat\beta}$ are invertible. Note that the fermionic tangent space  splits as $d_\alpha\rightarrow\{d_{\hat\alpha},\,d_{\alpha'}\}$, i.e. the 32 directions split into $n$ associated with the supersymmetries and the remaining $32-n$ directions. In the same way the cotangent space splits as $E^\alpha\rightarrow \{E^{\hat\alpha},\,E^{\alpha'}\}$. We can pick coordinates adapted to this split
\begin{equation}
\Theta^\mu\rightarrow\{\vartheta^{\hat\mu},\,\upsilon^{\mu'}\}\,,
\end{equation}
where the $n$ $\vartheta$ transform linearly under the supersymmetries while the $32-n$ $\upsilon$ do not. The frame superfields, or supervielbeins, then take the form
\begin{equation}
E^A=dx^mE_m{}^A+d\vartheta^{\hat\mu}E_{\hat\mu}{}^A+d\upsilon^{\mu'}E_{\mu'}{}^A\,,
\end{equation}
with the diagonal components $E_m{}^a$, $E_{\hat\mu}{}^{\hat\alpha}$ and $E_{\mu'}{}^{\alpha'}$ invertible. Note that objects with a single primed index, such as $E_m{}^{\alpha'}$ or $E_{\mu'}{}^{\hat\alpha}$ must be at least linear in $\upsilon$. This will be important below when we restrict to the sub-superspace spanned by $(x^m,\,\vartheta^{\hat\mu})$ by setting $\upsilon^{\mu'}=0$.

Under a superisometry transformation the frame $\{E^a,\,E^\alpha\}$ will in general transform by a Lorentz rotation, but this can always be canceled by a local Lorentz transformation so we will take
\begin{equation}
\delta E^A=0\,,
\label{eq:deltaEA}
\end{equation}
just as in the bosonic case. This leads to
\begin{equation}
0=\mathcal L_{\delta z}E^A=i_{\delta z}dE^A+dK_{(\epsilon)}^A=\nabla K_{(\epsilon)}^A+i_{K_{(\epsilon)}}T^A-E^Bi_{K_{(\epsilon)}}\Omega_B{}^A\,,
\label{eq:nablaKA}
\end{equation}
where $T^A$, the superspace torsion, was absent in the bosonic case. The condition that the background fields respect the isometries are
\begin{equation}
0=\delta\phi=i_{K_{(\epsilon)}}d\phi\,,\qquad 0=\delta H=di_{K_{(\epsilon)}}H\,,\qquad 0=\delta F^{(n)}=\mathcal L_{K_{(\epsilon)}}F^{(n)}\,,
\end{equation}
where $\phi$, $H$ and $F^{(n)}$ are the dilation, NSNS three-form and RR field strengths respectively. The same must be true for the torsion $T^A$ and Riemann curvature superfield $R^{AB}$. Making use of the invariance of the frame under the superisometries these conditions can be written in components as
\begin{align}
&K_{(\epsilon)}^Ad_A\phi=0\,,\qquad K_{(\epsilon)}^Dd_DH_{ABC}=0\,,\qquad K_{(\epsilon)}^Dd_DF^{(n)}_{A_1...A_n}=0\,,
\qquad K_{(\epsilon)}^Dd_DT_{BC}{}^A=0\,,
\nonumber\\
&K_{(\epsilon)}^Ed_ER_{CD}{}^{AB}=0\,.
\label{eq:field-cond}
\end{align}
We can write the condition on the NSNS-flux in a slightly different way by noting that the two-form potential $B$ must transform by a gauge transformation under the isometries which implies, as we saw in the bosonic case, that
\begin{equation}
i_KH=d\Lambda\,,
\end{equation}
for some one-form superfield $\Lambda$. This $\Lambda$ will enter the superisometry Noether current as will be seen below.

We will now prove that the $(10|n)$-dimensional sub-superspace spanned by the coordinates associated to isometries $(x^m,\,\vartheta^{\hat\mu})$ is a supercoset space. Calling this space $\mathcal M_{10|n}$ we have
\begin{equation}
\mathcal M_{10|n}=\frac{\mathcal G}{H}\,,
\end{equation}
where $\mathcal G$ is the superisometry (super)group whose Lie algebra we will determine. We therefore have the following sub-(super)spaces
\begin{equation}
\mathcal M_{10|0}\subset\mathcal M_{10|n}\subset\mathcal M_{10|32}\,,
\end{equation}
of the full superspace $\mathcal M_{10|32}$, where $\mathcal M_{10|0}$ is the bosonic symmetric space $\frac{G}{H}$. This fact may be intuitively obvious from the analogy with the bosonic case but the proof will nevertheless be instructive.

\subsection{The supercoset subspace and superisometry algebra}
We will first prove that certain superfields vanish when restricted to $\upsilon=0$. This will allow us to construct the supergeometry of this sub-superspace explicitly. We will then show that the same geometry follows from a supercoset construction with the superisometry Lie algebra taking a certain form.

Let us expand out the condition that the dilaton superfield respect the superisometries
\begin{equation}
0=K_{(\epsilon)}^Ad_A\phi=\epsilon^aK_a{}^Bd_B\phi+\epsilon^{ab}K_{ab}{}^Cd_C\phi+\epsilon^{\hat\alpha}K_{\hat\alpha}{}^Cd_C\phi\,.
\end{equation}
In particular we have
\begin{equation}
K_a{}^bd_b\phi=-K_a{}^{\hat\beta}d_{\hat\beta}\phi-K_a{}^{\beta'}d_{\beta'}\phi\,,\qquad
K_{\hat\alpha}{}^{\hat\beta}d_{\hat\beta}\phi
=
-K_{\hat\alpha}{}^{\gamma'}d_{\gamma'}\phi
-K_{\hat\alpha}{}^bd_b\phi\,.
\end{equation}
Restricting these equations to $\upsilon=0$ and using the fact that $K_b{}^{\beta'}|=0=K_{\hat\alpha}{}^{\gamma'}|$ and that $K_a{}^b$ and $K_{\hat\alpha}{}^{\hat\beta}$ are invertible we find
\begin{equation}
d_a\phi|=
-(K^{-1})_a{}^bK_b{}^{\hat\beta}d_{\hat\beta}\phi|
\,,\qquad
(K_{\hat\alpha}{}^{\hat\beta}-K_{\hat\alpha}{}^b(K^{-1})_b{}^cK_c{}^{\hat\beta})d_{\hat\beta}\phi|=0\,,
\end{equation}
where we have used the first equation in the last. Now we note that the matrix appearing in the last equation is invertible\footnote{An invertible bosonic matrix plus fermion terms is invertible.} so that finally we find
\begin{equation}
d_a\phi|=0\,,\qquad d_{\hat\alpha}\phi|=0\,.
\end{equation}
Applying the same calculation to the other background superfields the invariance conditions in eq. (\ref{eq:field-cond}) give
\begin{align}
&\nabla_aH_{bcd}|=0\,,\qquad\nabla_{\hat\alpha}H_{bcd}|=0\,,\qquad
\nabla_a\mathcal S|=0\,,\qquad\nabla_{\hat\alpha}\mathcal S|=0\,,\qquad
\nonumber\\
&
\nabla_{\hat\alpha}\chi^\beta|=0\,,\qquad
\nabla_{\hat\alpha}\psi_{ab}^\beta|=0\,,\qquad
\nabla_aR_{bc}{}^{de}|=0\,,\qquad\nabla_{\hat\alpha}R_{bc}{}^{de}|=0\,,
\label{eq:supercoset-cond1}
\end{align}
where we have combined the RR field strengths into the bispinor superfield $\mathcal S$ defined as
\begin{equation}
\mathcal S=e^\phi\left\{
\begin{array}{cc}
\frac12F^{(2)}_{ab}\Gamma^{ab}\Gamma_{11}+\frac{1}{4!}F^{(4)}_{abcd}\Gamma^{abcd}\\
-iF^{(1)}_a\gamma^a\sigma^2-\frac{1}{3!}F^{(3)}_{abc}\gamma^{abc}\sigma^1-\frac{i}{2\cdot5!}F^{(5)}_{abcde}\gamma^{abcde}\sigma^2
\end{array}
\right.
\,,
\label{eq:S}
\end{equation}
and used the fact that $F^{(n)}_{\alpha a_1\cdots a_{n-1}}\sim\chi$ and $T_{ab}{}^\alpha=\psi_{ab}^\alpha$ from the superspace constraints, where $\chi$ is the dilatino superfield and $\psi_{ab}$ is the gravitino field strength superfield. In principle there are also constraints on other components of the superfields, e.g. $H_{\alpha\beta c}$, but these are implied by the above equations since these involve all the basic superfields. 
Note that the conditions have been written in a Lorentz-covariant way involving the covariant derivative $\nabla$. Since the original equations involved the ordinary derivative however we also have
\begin{equation}
\Omega_{a[b}{}^eH_{cd]e}|=0\,,\qquad\Omega_{\hat\alpha[b}{}^eH_{cd]e}|=0\,,
\end{equation}
and similarly for the other fields, i.e. the spin connection acts trivially on them. These are the analog of the conditions found on $H$ in the bosonic case, eq. (\ref{eq:bosonic-H-cond}). By noting that we must also have\footnote{In principle we could have $d_{\alpha'}\phi|=\vartheta^{\hat\mu}X_{\hat\mu\alpha'}$ but the condition that $d_{\alpha'}\phi$ respect the superisometries implies $\nabla_ad_{\alpha'}\phi|=\nabla_{\hat\alpha}d_{\alpha'}\phi|=0$ which means that $X_{\hat\mu\alpha'}=0$.
} $d_{\alpha'}\phi|=0$ we find
\begin{equation}
d\phi|=0\,,\qquad\Rightarrow\qquad\chi_\alpha|\equiv d_\alpha\phi|=0\,,
\end{equation}
i.e. the dilatino superfield $\chi^\alpha$ vanishes when restricted to the $\upsilon=0$ subspace. Using this fact together with the superspace constraints (see the appendix of \cite{Wulff:2013kga}) the invariance conditions in eq. (\ref{eq:supercoset-cond1}) reduce to the following (here and in the following all fields are understood to be restricted to $\upsilon=0$):
\begin{itemize}
	\item[1.] Vanishing of the fermionic superfields
	\begin{equation}
\chi^\alpha=0\,,\qquad\psi_{ab}^\alpha=0\,.
\label{eq:inv-fermions}
	\end{equation}
	\item[2.] Constancy of the bosonic fields (i.e. invariance with respect to translations)
\begin{equation}
\nabla_a\phi=0\,,\qquad\nabla_aH_{bcd}=0\,,\qquad\nabla_a\mathcal S=0\,,\qquad\nabla_aR_{bc}{}^{de}=0\,.
\label{eq:inv-translations}
\end{equation}
\item[3.] Invariance with respect to rotational isometries
\begin{equation}
\Omega_{[a}{}^dH_{bc]d}=0\,,\qquad [\Omega^{ab}\Gamma_{ab},\mathcal S]=0\,,\qquad\Omega_{[a}{}^cR_{b]c}{}^{de}=0\,.
\label{eq:inv-rotations}
\end{equation}
\item[4.] Invariance with respect to supersymmetries
\begin{equation}
\left(\frac16H_{abc}\Gamma^{abc}\Gamma_{11}+\frac14\Gamma^a\mathcal S\Gamma_a\right)^\alpha{}_{\hat\beta}=0\,,\qquad
\left(G_{[a}G_{b]}-8R_{ab}{}^{cd}\Gamma_{cd}\right)^\alpha{}_{\hat\beta}=0\,.
\label{eq:inv-susy}
\end{equation}
\end{itemize}
The matrices $G_a$ and $G_{ab}$ come from the dimension 1 torsion and curvature superspace constraints (recall that the type IIB expressions are obtained by (\ref{eq:IIAtoIIB}))
\begin{align}
T_{a\beta}{}^\gamma=&-\frac18(H_{abc}\Gamma^{bc}\Gamma_{11}+\mathcal S\Gamma_a)^\gamma{}_\beta\equiv-\frac18(G_a)^\gamma{}_\beta\,,
\label{eq:Talpha-const}
\\
R_{\alpha\beta}{}^{cd}=&\frac{i}{2}H^{cde}\,(\Gamma_e\Gamma_{11})_{\alpha\beta}-\frac{i}{4}(\Gamma^{[c}\mathcal S\Gamma^{d]})_{\alpha\beta}\equiv\frac{i}{4}(G^{cd})_{\alpha\beta}
=-\frac{i}{4}(\Gamma^cG^d)_{(\alpha\beta)}
\,.
\label{eq:Ralphabeta-const}
\end{align}
The conditions coming from invariance under the supersymmetries, eq. (\ref{eq:inv-susy}), are the dilatino equation and the integrability condition for the Killing spinor equation, see \cite{Wulff:2014kja}.

We are now ready to find explicitly the form of the supergeometry of this sub-superspace.

In general the full supergeometry can be found by solving a set of first order ODE's for the $\Theta$-dependence, with initial condition given by the bosonic geometry. These equations were written for the full type II superspace in \cite{Wulff:2013kga}, where they were also solved to fourth order in $\Theta$. Restricting these equations to the $\upsilon=0$ subspace and using the fact that the dilatino and gravitino field strength superfields vanish this system of equations simplifies enormously and we are left with (where we've taken $\vartheta\rightarrow t\vartheta$)
\begin{align}
\frac{d}{dt}E^a=-iE\Gamma^a\vartheta\,,\qquad
\frac{d}{dt}E^{\hat\alpha}=([d-\frac14\Omega^{ab}\Gamma_{ab}+\frac18E^aG_a]\vartheta)^{\hat\alpha}\,,\qquad
\frac{d}{dt}\Omega^{ab}=-\frac{i}{4}\vartheta G^{ab}E\,,
\end{align}
while $E^{\alpha'}|=0$ which implies the consistency conditions
\begin{equation}
\Omega^{ab}(\Gamma_{ab})^{\alpha'}{}_{\hat\beta}=0\,,\qquad (G_a)^{\alpha'}{}_{\hat\beta}=0\,.
\label{eq:consistency-cond}
\end{equation}
All other superfields, $\phi,\,H_{abc},\,\mathcal S,\,R^{ab}$ are independent of $\vartheta$. In fact they must be constant since the bosonic geometry is a symmetric space, eq. (\ref{eq:inv-translations}). 

The equations for the $\vartheta$-dependence of the supervielbeins and spin connection are easily solved, but instead of doing this we will show that the same equations follow from a supercoset construction. Let $\mathcal G$ be a supergroup to be specified and consider the coset
\begin{equation}
\frac{\mathcal G}{H}\,.
\end{equation}
The Maurer-Cartan form is again
\begin{equation}
J=g^{-1}dg\,,\qquad g\in\mathcal G\,.
\end{equation}
We take a coset representative of the form
\begin{equation}
g=e^{x^aP_a}e^{\vartheta^{\hat\alpha}Q_{\hat\alpha}}\,.
\end{equation}
At $\vartheta=0$ we should get our ordinary bosonic symmetric space $\frac{G}{H}$ which implies that the bosonic subgroup of $\mathcal G$ should be $G$. Writing
\begin{equation}
J=\frac12\Omega^{ab}M_{ab}+E^aP_a+E^{\hat\alpha}Q_{\hat\alpha}\,,
\end{equation}
we can determine the $\vartheta$-expansion of the supervielbeins and spin connection by taking $\vartheta\rightarrow t\vartheta$ in the coset representative and using the Maurer-Cartan equation to find linear ODE's that determine the $t$-, i.e. $\vartheta$-dependence. We find the equations
\begin{equation}
\frac12\frac{d}{dt}\Omega^{ab}M_{ab}+\frac{d}{dt}E^aP_a+\frac{d}{dt}E^{\hat\alpha}Q_{\hat\alpha}=d\vartheta^{\hat\alpha}Q_{\hat\alpha}+[J(t),\vartheta^{\hat\alpha} Q_{\hat\alpha}]\,.
\end{equation}
These equations precisely coincide with the ones found above from the consideration of the superspace constraints provided that the bosonic commutators are the same as found before, eq. (\ref{eq:bosonic-algebra}), while the commutators involving $Q_{\hat\alpha}$ in the Lie algebra of $\mathcal G$ should take the form
\begin{align}
[P_a,Q_{\hat\alpha}]=&\frac18(QG_a)_{\hat\alpha}\,,\qquad[M_{ab},Q_{\hat\alpha}]=-\frac12(Q\Gamma_{ab})_{\hat\alpha}\,,
\nonumber\\
\{Q_{\hat\alpha},Q_{\hat\beta}\}=&i(\Gamma^a)_{\hat\alpha\hat\beta}\,P_a-\frac{i}{8}(G^{ab})_{\hat\alpha\hat\beta}\,M_{ab}\,.
\label{eq:Q-algebra}
\end{align}
Note that the matrices $G_a$ and $G_{ab}$ defined in terms of the fluxes in eqs. (\ref{eq:Talpha-const}) and (\ref{eq:Ralphabeta-const}) are indeed constant as required. To show that this is a (super) Lie algebra we need to verify the Jacobi identities. It is not hard to show, using the invariance conditions eqs. (\ref{eq:inv-rotations}) and (\ref{eq:inv-susy}) together with the consistency conditions in eq. (\ref{eq:consistency-cond}) and the definitions of $G_a$ and $G_{ab}$ in eqs. (\ref{eq:Talpha-const}) and (\ref{eq:Ralphabeta-const}),  that they are indeed satisfied. The only somewhat non-trivial cases are the $QPP$ and $QQQ$ Jacobi identities. The former requires multiplying the Killing spinor integrability condition in eq. (\ref{eq:inv-susy}) by a gamma-matrix, symmetrizing in the spinor indices and noting that the Bianchi identity for the curvature then implies that
\begin{equation}
(G_{[ab}G_{c]})_{(\hat\alpha\hat\beta)}=0\,.
\end{equation}
To verify the $QQQ$ Jacobi identity one must use the symmetry properties of $\mathcal S$, i.e. $\mathcal S_{(\alpha\beta)}=0$ as follows from eq. (\ref{eq:S}), and the basic Fierz identity
\begin{equation}
\Gamma^a_{(\alpha\beta}(\Gamma_a\Gamma_{11})_{\gamma\delta)}=0\,.
\label{eq:Fierz}
\end{equation}

We can derive further conditions from the requirement that there exist an invariant bilinear form, which we denote $\mathrm{Str}(\cdots)$, on the Lie superalgebra such that
\begin{equation}
\mathrm{Str}(P_aP_b)=c_{[a]}\eta_{ab}\,,
\end{equation}
where $c_{[a]}$ are constants that can take a different value for each irreducible factor in the geometry. They are related to the relative radius of curvature of the different factors in the geometry. For the backgrounds listed in table \ref{table:1} and \ref{table:2} $c_{[a]}$ takes values $1$ or $2$ in our conventions.

Taking the trace of the $QQ$-commutator in eq. (\ref{eq:Q-algebra}) with $P_a$ we get\footnote{We see that at least some of the fluxes must be non-zero for this to be possible since otherwise $G_a$ vanishes. In that case we would have flat space and we could solve the theory directly.}
\begin{equation}
ic_{[a]}(C\Gamma_a)_{\hat\alpha\hat\beta}
=
\mathrm{Str}([P_a,Q_{\hat\alpha}]Q_{\hat\beta})
=
\frac{i}{8}(\mathcal KG_a)_{\hat\beta\hat\alpha}\,,
\label{eq:Kalphabeta-cond}
\end{equation}
where $C$ is the charge-conjugation matrix (which we leave implicit when there is no risk of confusion, e.g. in eq. (\ref{eq:Q-algebra})) and
\begin{equation}
\mathcal K_{\hat\alpha\hat\beta}=i\mathrm{Str}(Q_{\hat\alpha}Q_{\hat\beta})
\end{equation}
is the anti-symmetric invariant metric in the fermionic sector. We see from eq. (\ref{eq:Kalphabeta-cond}) that $\mathcal K_{\hat\alpha\hat\beta}$ must be non-degenerate and therefore it follows that there exists a matrix $\widehat{\mathcal K}^{\hat\alpha\hat\beta}$ such that
\begin{equation}
\widehat{\mathcal K}\mathcal K=\mathcal P\,,
\end{equation}
where $\mathcal P$ is the projector whose only non-zero entries are $\mathcal P^{\hat\alpha}{}_{\hat\beta}=\delta^{\hat\alpha}_{\hat\beta}$, i.e. $\mathcal P$ projects on the fermionic directions associated with supersymmetries. It will be somewhat more convenient to work instead, as in \cite{Wulff:2014kja}, with the object
\begin{equation}
\hat{\mathcal S}\equiv8\widehat{\mathcal K}C
\end{equation}
which has the same placement of spinor indices as the gamma matrices. In terms of this object we then have the relation
\begin{equation}
(G_a)^\alpha{}_{\hat\beta}=c_{[a]}(\hat{\mathcal S}\Gamma_a)^\alpha{}_{\hat\beta}\,.
\label{eq:Ga-hatS}
\end{equation}
From the definition of $G_a$ in eq. (\ref{eq:Talpha-const}) we see that $\hat{\mathcal S}$ provides, in a sense, an analog of the RR bispinor $\mathcal S$ for cases with non-zero NSNS flux. In the case of zero NSNS flux we find that we must take
\begin{equation}
H_{abc}=0:\qquad\hat{\mathcal S}=\mathcal S\qquad c_{[a]}=1\,.
\end{equation}
It is important to note that this conclusion only holds if there is some non-zero amount of supersymmetry preserved since otherwise $\hat{\mathcal S}$ does not exist but $\mathcal S$ of eq. (\ref{eq:S}) certainly does.

Similarly we find by taking the trace of the $QQ$-commutator with $M_{ab}$ that
\begin{equation}
(G_{ab})_{\hat\alpha\hat\beta}
=
-\frac{1}{c_{[a]}}R_{ab}{}^{cd}(\mathcal K\Gamma_{cd})_{\hat\alpha\hat\beta}
=
-c_{[a]}(C\Gamma_{[a}\hat{\mathcal S}\Gamma_{b]})_{\hat\alpha\hat\beta}\,.
\label{eq:Gab-K}
\end{equation}
In the last step we made use of the supersymmetry conditions in eq. (\ref{eq:inv-susy}). Using these relations the commutators involving $Q$ take the form
\begin{align}
[P_a,Q_{\hat\alpha}]=&\frac{c_{[a]}}{8}(Q\hat{\mathcal S}\Gamma_a)_{\hat\alpha}\,,\qquad[M_{ab},Q_{\hat\alpha}]=-\frac12(Q\Gamma_{ab})_{\hat\alpha}\,,
\nonumber\\
\{Q_{\hat\alpha},Q_{\hat\beta}\}=&i(\Gamma^a)_{\hat\alpha\hat\beta}\,P_a+\frac{ic_{[a]}}{8}(\Gamma^a\hat{\mathcal S}\Gamma^b)_{\hat\alpha\hat\beta}\,M_{ab}\,.
\label{eq:Q-algebra2}
\end{align}

Notice that the relation (\ref{eq:Gab-K}) implies
\begin{equation}
0=(G_{ab'})_{\hat\alpha\hat\beta}=-(\Gamma_aG_{b'})_{(\hat\alpha\hat\beta)}\,,
\end{equation}
where $b'$ denotes any flat direction. Since this must hold for all $a$ there is generically no solution except
\begin{equation}
(G_{a'})^\alpha{}_{\hat\beta}=0\,.
\end{equation}
From eqs. (\ref{eq:Ga-hatS}) and (\ref{eq:Kalphabeta-cond}) we then find the conditions (we insist that $c_{[a']}\neq0$)
\begin{equation}
(\hat{\mathcal S}\Gamma_{a'})^\alpha{}_{\hat\beta}=0\,,\qquad (\Gamma_{a'})_{\hat\alpha\hat\beta}=0\,.
\label{eq:aprime-rel}
\end{equation}
These conditions imply that the translations in the flat directions $P_{a'}$ completely decouple from the algebra in eq. (\ref{eq:Q-algebra2}) and therefore just generate $U(1)$ factors.

\subsection{Special case: (semi)symmetric superspace}\label{sec:semisymmetric}
The definition of a (bosonic) symmetric space involves a $\mathbbm Z_2$-automorphism, or involution, of the Lie algebra $\mathfrak g$ which acts as
\begin{equation}
P_a\rightarrow-P_a\,,\qquad M_{ab}\rightarrow M_{ab}\,.
\end{equation}
In general this automorphism will not extend to the full superisometry algebra and $\frac{\mathcal G}{H}$ will just be a homogeneous superspace. In the case when it extends to a $\mathbbm Z_2$-automorphism of the superalgebra we get a symmetric superspace. More generally, if it extends to a automorphism which is not $\mathbbm Z_2$, the corresponding supercoset space is called a semisymmetric superspace \cite{Serganova:1983vp}. In that case it must act on the fermionic generators as
\begin{equation}
Q_{\hat\alpha}\rightarrow Q_{\hat\beta}\Sigma^{\hat\beta}{}_{\hat\alpha}\,,
\end{equation}
where the matrix $\Sigma$ satisfies the following relations, needed to preserve the form of the commutators in eq. (\ref{eq:Q-algebra}),
\begin{align}
&(\Sigma G_a\Sigma^{-1})^{\hat\beta}{}_{\hat\alpha}=-(G_a)^{\hat\beta}{}_{\hat\alpha}\,,\quad
(\Sigma\Gamma^{cd}\Sigma^{-1})^{\hat\beta}{}_{\hat\alpha}R_{cd}{}^{ab}=(\Gamma^{cd})^{\hat\beta}{}_{\hat\alpha}R_{cd}{}^{ab}\,,\quad
(\Sigma^{\mathrm T}C\Gamma^a\Sigma)_{\hat\alpha\hat\beta}=-(C\Gamma^a)_{\hat\alpha\hat\beta}\,,
\nonumber\\
&(\Sigma^{\mathrm T}G^{ab}\Sigma)_{\hat\alpha\hat\beta}=(G^{ab})_{\hat\alpha\hat\beta}\,.
\end{align}
Using instead the form of the algebra in eq. (\ref{eq:Q-algebra2}) and simplifying we see that these conditions can be reduced to the two conditions
\begin{equation}
(\Sigma^{\mathrm T}\mathcal K\Sigma)_{\hat\alpha\hat\beta}=\mathcal K_{\hat\alpha\hat\beta}\,,\qquad
(\Sigma^{\mathrm T}C\Gamma^a\Sigma)_{\hat\alpha\hat\beta}=-(C\Gamma^a)_{\hat\alpha\hat\beta}\,.
%
\label{eq:aut-cond}
\end{equation}

\subsubsection{$H_{abc}=0$ case}
Consider the case when the NSNS flux vanishes, i.e. $H_{abc}=0$. As we have seen we have in this case $\mathcal S=\hat{\mathcal S}=\mathcal P\hat{\mathcal S}$ (as already remarked this is only true if the background preserves some amount of supersymmetry, but this is precisely the case that is relevant here). Since furthermore $[\mathcal S,\Gamma_{11}]=0$ (or $\{\mathcal S,\sigma^3\}=0$ for type IIB), as follows from eq. (\ref{eq:S}) , we conclude that $\Gamma_{11}$ commutes with $\mathcal P$ or equivalently
\begin{equation}
(\Gamma_{11})^{\alpha'}{}_{\hat\beta}=0\,,\qquad(\Gamma_{11})^{\hat\alpha}{}_{\beta'}=0\,.
\end{equation}
Taking
\begin{equation}
\Sigma^{\hat\alpha}{}_{\hat\beta}=i(\Gamma_{11})^{\hat\alpha}{}_{\hat\beta}
\end{equation}
it is easy to see that the conditions in eq. (\ref{eq:aut-cond}) are satisfied. Furthermore we have $\Sigma^2=-1$ which means that we have a $\mathbbm Z_4$-automorphism of the superisometry algebra. Therefore, when there is no NSNS flux, the supercoset sub-superspace is a semisymmetric superspace.

\subsubsection{$H_{abc}\neq0$ case}
When there is non-zero NSNS flux the supercoset subspace may or may not be a semisymmetric superspace. Let us consider the examples which are known to exhibit integrability. For $AdS_3\times S^3\times S^3\times S^1$ one can show that
\begin{equation}
\Sigma=i\Gamma_{11'}\equiv i(q\Gamma^9+\hat q\Gamma_{11})\,,
\end{equation}
where $q^2+\hat q^2=1$ and $q$ controls the amount of NSNS flux while $\hat q$ controls the amount of RR flux, satisfies the conditions in eq. (\ref{eq:aut-cond}). In particular it is easy to see from the form of the projector $\mathcal P$ and the matrix $\hat{\mathcal S}$ given in \cite{Wulff:2014kja} that
\begin{equation}
[\Gamma_{11'},\mathcal P]=0\,,\quad[\Gamma_{11'},\hat{\mathcal S}]=0\,,\quad\{\Gamma_{11'},\Gamma^a\}=0\,\,\,(a\neq9)
\,.
\label{eq:gamma11-rel}
\end{equation}
This, together with eq. (\ref{eq:aprime-rel}), is enough to verify the conditions in eq. (\ref{eq:aut-cond}), and since $\Sigma^2=-1$ we again have a $\mathbbm Z_4$-automorphism. 

Similarly for $AdS_2\times S^2\times S^2\times T^4$ we find, using the expressions for $\mathcal P$ and $\hat{\mathcal S}$ given in \cite{Wulff:2014kja} that $\Sigma=i\Gamma_{11'}$ satisfies the same conditions as in the $AdS_3\times S^3\times S^3\times S^1$ case, and again gives rise to a $\mathbbm Z_4$-automorphism.

Considering for example $AdS_3\times S^2\times S^3\times T^2$ we find that we would need to take $\Sigma=i\Gamma^9$. This satisfies the same conditions as above \emph{except} that now $\Gamma^9$ anti-commutes, rather than commutes, with $\hat{\mathcal S}$. This extra sign means that it is not possible to satisfy the conditions in eq. (\ref{eq:aut-cond}) and so the supercoset subspace in this case is not a semisymmetric superspace. The same turns out to be true for the other examples with a mix of two and three-dimensional factors in the geometry in table \ref{table:2}. Since the string was shown to be integrable also for these backgrounds in \cite{Wulff:2014kja} the existence of a semisymmetric sub-superspace, or a $\mathbbm Z_4$-automorphism of the isometry algebra, cannot be a necessary condition for (classical) integrability.

\subsection{Killing vectors and Killing spinors}
We can now construct the Killing vectors and Killing spinors valued in the Lie algebra as we did for the Killing vectors in the bosonic case (recall that $\widehat{\mathcal K}\mathcal K=\mathcal P$)\footnote{Note that with this choice the supervielbeins transform by a local Lorentz transformation $\delta E^a=E^bl_b{}^a$, $\delta E^\alpha=-\frac14(\Gamma^{ab}E)^\alpha l_{ab}$ which must be included in eqs. (\ref{eq:deltaEA}) and (\ref{eq:nablaKA}) just as in the bosonic case eq. (\ref{eq:nablaKa}).}
\begin{equation}
K^a=gP^ag^{-1}\,,\qquad\xi^{\hat\alpha}=-i\widehat{\mathcal K}^{\hat\alpha\hat\beta}gQ_{\hat\beta}g^{-1}\,.
\end{equation}
Making use of the algebra in eq. (\ref{eq:Q-algebra}) and the definition of the Maurer-Cartan form we find
\begin{equation}
\nabla K^a=-i\xi\Gamma^aE-\frac12E^bR_b{}^{acd}gM_{cd}g^{-1}
\end{equation}
and
\begin{equation}
\nabla\xi^{\hat\alpha}
=
-\frac18E^a\,(G_a\xi)^{\hat\alpha}
+\frac18(G_aE)^{\hat\alpha}\,K^a
+\frac18(\Gamma^{ab}E)^{\hat\alpha}\,R_{ab}{}^{cd}gM_{cd}g^{-1}\,.
\label{eq:nabla-xi}
\end{equation}
These equations precisely coincide with the superspace Killing equations in eq. (\ref{eq:nablaKA}), with the local Lorentz transformation $l^{ab}$ added, restricted to $\upsilon=0$ (with the constant parameters $\epsilon$ replaced by generators of the Lie algebra) upon the identifications
\begin{equation}
\nabla^aK^b=K^C\Omega_C{}^{ab}+l^{ab}=-\frac12R^{abcd}gM_{cd}g^{-1}\,,\qquad K^{\hat\alpha}=\xi^{\hat\alpha}\,.
\end{equation}
Using their definition and the algebra in eq. (\ref{eq:Q-algebra2}) the Killing vectors and Killing spinors constructed in this way are found to satisfy the following commutation relations (suppressing the spinor indices)
\begin{align}
&[K_a,K_b]=\nabla_aK_b
\,,\qquad
[K_c,\nabla_aK_b]=R_{abc}{}^dK_d
\,,\qquad
[K_a,\xi]=-\frac{c_{[a]}}{8}\hat{\mathcal S}\Gamma_a\xi=\nabla_a\xi
\,,
\nonumber\\
&[\nabla_aK_b,\xi]=-\frac14R_{ab}{}^{cd}\Gamma_{cd}\xi
\,,\qquad
\{\xi,\xi\}=-\frac{i}{64}(\hat{\mathcal S}\Gamma^a\hat{\mathcal S}C)\,K_a+\frac{i}{32c_{[\hat a]}}(\Gamma^{ab}\hat{\mathcal S}C)\,\nabla_aK_b\,.
\end{align}
This form of the algebra of Killing vectors and Killing spinors (with $\xi$ rescaled by $\sqrt8$) was postulated already in \cite{Wulff:2014kja} based on experience with specific cases. Here we have derived it from first principles.

\subsection{Superisometry Noether current}
Here we will underline the quantities defined in the full $10|32$-dimensional superspace to distinguish them from the corresponding supercoset quantities, e.g. $\u E^a=E^a(x,\vartheta,\upsilon)$ and $E^a=\u E^a|_{\upsilon=0}$. The superisometry Noether current of the string takes the simple form
\begin{equation}
\mathcal J=\u E^a\u K_a+*\u\Lambda\,,\qquad\mbox{where}\qquad d\u\Lambda=i_{\u K}\u H\,.
\label{eq:calJ}
\end{equation}
It is easily seen to be conserved
\begin{align}
d*\mathcal J=&
d(*\u E^a\u K_a)+i_{\u K}\u H
=
\u\nabla*\u E^a\u K_a
-*\u E^ai_{\u K}\u T^b\eta_{ab}
+i_{\u K}\u H
\nonumber\\
=&
\u K_a\big(\u\nabla*\u E^a-\frac{i}{2}\u E\Gamma^a\Gamma_{11}\u E+\frac12\u E^c\u E^b\u H^a{}_{bc}\big)
-i\u\xi\big(*\u E^a\,\Gamma_a\u E-\u E^a\,\Gamma_a\Gamma_{11}\u E\big)\,,
\end{align}
where we used eq. (\ref{eq:nablaKA}) and the superspace constraint on the NSNS three-form $\u H$ and torsion $\u T^a$. The first term is proportional to the bosonic equations of motion while the second is proportional to the fermionic equations of motion so that indeed $d*\mathcal J=0$ on-shell.

We can expand $\mathcal J$ in the generators of the superisometry Lie algebra (or rather in the Killing vectors and Killing spinors) as
\begin{equation}
\mathcal J=g(\mathcal J^aP_a+\frac12\mathcal J^{ab}M_{ab}+\mathcal J^{\hat\alpha}Q_{\hat\alpha})g^{-1}\,.
\label{eq:Jsplit}
\end{equation}
The conservation of $\mathcal J$ then implies the following equations for the components
\begin{align}
\nabla*\mathcal J^a+E^b*\mathcal J_b{}^a+iE\Gamma^a*\mathcal J=&0\,,
\label{eq:J-cons-comp1}
\\
\nabla*\mathcal J^{ab}+E^c*\mathcal J^dR_{cd}{}^{ab}-\frac{i}{4}EG^{ab}*\mathcal J=&0\,,
\label{eq:J-cons-comp2}
\\
\nabla*\mathcal J^{\hat\alpha}
-\frac18E^a\,(G_a*\mathcal J)^{\hat\alpha}
-\frac18*\mathcal J^a\,(G_aE)^{\hat\alpha}
+\frac14*\mathcal J^{ab}\,(\Gamma_{ab}E)^{\hat\alpha}
=&0\,.
\label{eq:J-cons-comp3}
\end{align}
These equations will be useful when constructing the Lax connection for the string in the next section.

In appendix \ref{app:J} $\mathcal J$ is computed up to quadratic order in the non-coset fermions $\upsilon$.

\section{Integrability for the superstring}\label{sec:integrability}
We will first treat the case of zero NSNS flux for which we will find a general Lax connection up to quadratic order in the non-coset fermions $\upsilon$. For the case of non-zero NSNS flux we have not found a general construction so we will concentrate on the known examples in table \ref{table:2}.

\subsection{General Lax connection: $H_{abc}=0$ case}
We will simply postulate a Lax connection and then show that its flatness implies the equations of motion to quadratic order in $\upsilon$. We take
\begin{equation}
L=
\frac12\Omega^{ab}M_{ab}
+\frac{(1+\alpha)\beta}{2}*\mathcal J^{ab}M_{ab}
+\frac{\beta^2}{2}\mathcal J^{ab}M_{ab}
+(1+\alpha)E^aP_a
+\beta*\mathcal J^aP_a
+QVE
+\beta QV^\dagger*\mathcal J'\,,
\label{eq:super-Lax}
\end{equation}
where all dependence on the non-coset fermions $\upsilon$ comes through the components of the superisometry Noether current $\mathcal J$. Here we have defined
\begin{equation}
\mathcal J'^{\hat\alpha}=\mathcal J^{\hat\alpha}+\frac12\Gamma_{11}*E
\label{eq:Jprime}
\end{equation}
and
\begin{equation}
V=\frac{1}{\sqrt2}(\sqrt{2+\alpha}-\sqrt{\alpha}\,\Gamma_{11})\,,\qquad V^\dagger\equiv-CV^{\mathrm T}C =\frac{1}{\sqrt2}(\sqrt{2+\alpha}+\sqrt{\alpha}\,\Gamma_{11})\,.
\label{eq:V-RR}
\end{equation}
The parameters $\alpha$ and $\beta$ are related by the same equation as before
\begin{equation}
\beta^2=(1+\alpha)^2-1\,,
\end{equation}
which implies that the matrix $V$ satisfies the relations
\begin{equation}
VV^\dagger=1\,,\qquad V^2=1+\alpha-\beta\Gamma_{11}\,.
\label{eq:Vids}
\end{equation}
The $\mathbbm Z_4$-automorphism acts as\footnote{A convenient choice that is often made is
$$
\alpha=\frac{2\rm x^2}{1-\rm x^2}\,,\qquad\beta=\frac{2\rm x}{1-\rm x^2}
$$
so that the $\mathbbm Z_4$-automorphism acts by inversion of the spectral parameter $\rm x\rightarrow\frac{1}{\rm x}$.
}
\begin{equation}
\alpha\rightarrow-2-\alpha\,,\qquad\beta\rightarrow-\beta\,,\qquad V\rightarrow-i\Gamma_{11}V\,.
\end{equation}
When restricted to $\upsilon=0$ this Lax connection reduces to the standard Lax connection for a $\mathbbm Z_4$-symmetric supercoset sigma-model.\footnote{This is often written in terms of the components of the Maurer-Cartan form split according to the $\mathbbm Z_4$-grading:
$$
J_0=\frac12\Omega^{ab}M_{ab}\,,\quad J_2=E^aP_a\,,\quad J_1=(\frac12(1+\Gamma_{11})E)^\alpha Q_\alpha\,,\quad J_3=(\frac12(1-\Gamma_{11})E)^\alpha Q_\alpha\,.
$$
}

Using the conservation equations for the components of the Noether current in eqs. (\ref{eq:J-cons-comp1})--(\ref{eq:J-cons-comp3}), the superspace constraints for the supercoset subspace (with $H_{abc}=0$)
\begin{equation}
T^a=-\frac{i}{2}E\Gamma^aE\,,\quad
T^\alpha=\frac18E^a\,(\mathcal S\Gamma_aE)^\alpha\,,\quad
R^{ab}=-\frac{i}{8}E\Gamma^a\mathcal S\Gamma^bE+\frac12E^dE^cR_{cd}{}^{ab}\,,
\label{eq:supercoset-constr}
\end{equation}
and the superisometry algebra in eqs. (\ref{eq:bosonic-algebra}) and (\ref{eq:Q-algebra2}) we find for the terms in the curvature of $L$ proportional to $P_a$
\begin{align}
\frac{1}{\beta^2}[dL-LL]_{P_a}=&
\beta*(\mathcal J^b-E^b)\mathcal J_b{}^a
-(1+\alpha)(\mathcal J^b-E^b)\mathcal J_b{}^a
-\frac{i}{2}\mathcal J'V^2\Gamma^a\mathcal J'\,.
\end{align}
Note that we have used the identities satisfied by $V$ in eq. (\ref{eq:Vids}) to simplify the expression. The $\upsilon$-expansion of $\mathcal J$ in appendix \ref{app:J} shows that $\mathcal J^{ab}$ is of order $\upsilon^2$, and since $\mathcal J^a=E^a+\mathcal O(\upsilon)$ the first two terms are of order $\upsilon^3$. For the last term we observe that at the linear order in $\upsilon$ we have $\mathcal J'^{\hat\alpha}=A+\Gamma_{11}*A$ for a certain one-form $A$, which is easily seen to imply that the $\upsilon^2$ contribution from the last term vanishes. We therefore have
\begin{equation}
[dL-LL]_{P_a}=\mathcal O(\upsilon^3)\,.
\end{equation}
Next we turn to the components of the curvature proportional to $Q_{\hat\alpha}$. After a little bit of simplification we find
\begin{align}
\frac{1}{\beta^2}V^\dagger[dL-LL]_Q=&
-\frac18\mathcal J^a\,\mathcal S\Gamma_a\mathcal J'
+\frac18E^a\,\mathcal S\Gamma_a\Gamma_{11}*\mathcal J'
-\frac14\mathcal J^{ab}\,\Gamma_{ab}E
+\frac14*\mathcal J^{ab}\,\Gamma_{ab}\Gamma_{11}E
\nonumber\\
&{}
-\frac14\big((1+\alpha)*\mathcal J^{ab}+\beta\mathcal J^{ab}\big)\Gamma_{ab}(V^\dagger)^2*\mathcal J'\,.
\end{align}
The last term is clearly of cubic order in $\upsilon$ and we can forget about it. The remaining terms cancel to order $\upsilon^2$ due to the identity in eq. (\ref{eq:Jid1}) derived in the appendix.

Finally we have the terms in the curvature proportional to $M_{ab}$. They take the form
\begin{align}
\frac{1}{\beta^2}[dL-LL]_{M_{ab}}=&
\frac12\nabla\mathcal J^{ab}
+\frac14E^cE^dR_{cd}{}^{ab}
-\frac14\mathcal J^c\mathcal J^dR_{cd}{}^{ab}
-\frac{i}{8}E\Gamma^{[a}\mathcal S\Gamma_{11}\Gamma^{b]}*\mathcal J'
\nonumber\\
&{}
-\frac{i}{16}\mathcal J'\Gamma^a\mathcal S\Gamma^b\mathcal J'
-\frac12\mathcal J^{ac}\mathcal J_c{}^b\,.
\end{align}
The last term is quartic in $\upsilon$ and can be dropped. The remaining terms cancel to order $\upsilon^2$ due to the identity in eq. (\ref{eq:Jid2}) derived in the appendix. This shows that the Lax connection in eq. (\ref{eq:super-Lax}) is indeed flat, i.e.
\begin{equation}
dL-LL=\mathcal O(\upsilon^3)\,,
\end{equation}
when evaluated on the equations of motion. We also need to show that flatness implies the equations of motion. Going through the same calculation again but retaining terms proportional to the equations of motion we find that flatness of the Lax connection implies the conservation equations (\ref{eq:J-cons-comp1})--(\ref{eq:J-cons-comp3}) together with the following equation which arises in deriving the identity in eq. (\ref{eq:Jid2}) (here we are using the fact that the conservation equations imply the supercoset equations of motion when evaluated at $\upsilon=0$)
\begin{align}
\upsilon\Gamma^{cd}\Gamma_{11}
\Big(
*(E^e-iE\Gamma^e\upsilon)\,\Gamma_e(E+\mathcal D\upsilon)
-(E^e-iE\Gamma^e\upsilon)\,\Gamma_e\Gamma_{11}(E+\mathcal D\upsilon)
\Big)
\,R_{cd}{}^{ab}
=&{}\mathcal O(\upsilon^3)\,.
\label{eq:missing-eom}
\end{align}
Using the expansion of $\mathcal J$ given in the appendix we find that at the linear order in $\upsilon$ the conservation equations (\ref{eq:J-cons-comp1})--(\ref{eq:J-cons-comp3}) become, after using the lowest order equation, the supercoset torsion constraints eq. (\ref{eq:supercoset-constr}) and the Fierz identity eq. (\ref{eq:Fierz}),
\begin{align}
\nabla*(E^a-iE\Gamma^a\upsilon)
+\frac{i}{4}*E_b\,\upsilon\Gamma^{[b}\mathcal S\Gamma^{a]}E
-\frac{i}{2}E\Gamma^a\Gamma_{11}E
-iE\Gamma^a\Gamma_{11}\mathcal D\upsilon
=&\,\mathcal O(\upsilon^2)
\\
*(E^a-iE\Gamma^a\upsilon)\,(\mathcal S\Gamma_a(E+\mathcal D\upsilon))
-(E^a-iE\Gamma^a\upsilon)\,(\mathcal S\Gamma_a\Gamma_{11}(E+\mathcal D\upsilon))
=&\,\mathcal O(\upsilon^2)\,.
\end{align}
Using the expansion of the supervielbeins given in appendix \ref{app:J} it is easy to see that the first equation coincides with the expansion of the bosonic equation of motion in eq. (\ref{eq:eom}) to linear order in $\upsilon$ (note that the second term comes from expanding the spin connection). Similarly the second equation is precisely the fermionic equation of motion to linear order in $\upsilon$ projected onto the $n$ supersymmetry directions, or more precisely, it is
\begin{equation}
\Psi\mathcal P=0\,,
\end{equation}
where $\Psi=0$ is the fermionic equation of motion in eq. (\ref{eq:eom}) and $\mathcal P$ is the supersymmetry projector introduced previously. The conservation equations therefore do not give all the equations of motion but just, as should be expected, the ones for the coordinates associated with (super)isometries. But recall the extra condition we found, eq. (\ref{eq:missing-eom}), which takes the form (modulo higher order terms in $\upsilon$)
\begin{align}
\Psi\Gamma^{cd}\Gamma_{11}\upsilon\,R_{cd}{}^{ab}=0\,.
\end{align}
Using the fact that $(\Gamma^{cd}\Gamma_{11})^{\hat\alpha}{}_{\beta'}\,R_{cd}{}^{ab}=0$ this implies precisely the missing components of the fermionic equations of motion, i.e. the ones corresponding to the $32-n$ non-supersymmetric directions
\begin{equation}
\Psi(1-\mathcal P)=0\,.
\end{equation}
Therefore we conclude that the flatness of the Lax connection in eq. (\ref{eq:super-Lax}) is indeed equivalent to the string equations of motion (at least to quadratic order in $\upsilon$) which concludes the proof of the (classical) integrability.

It is important to note that in the above proof of the integrability we assumed that there is some non-zero amount of supersymmetry preserved by the background. This is because we used the relation $\mathcal S=\hat{\mathcal S}$ and in particular $\mathcal S=\mathcal P\mathcal S$ which, as we remarked earlier, only follow in the presence of supersymmetry. Computing the curvature of the Lax connection in eq. (\ref{eq:super-Lax}) in the non-supersymmetric case, in which case the $Q$-term is absent and the vielbeins and spin connection are just the bosonic ones, we find that the $P_a$-terms cancel as before but we are left with and $M_{ab}$-term coming from $\nabla\mathcal J^{ab}$ 
\begin{align}
dL-LL
\propto
E^cE^d\,\upsilon\Gamma_c\Gamma^{ab}\mathcal S\Gamma_d\upsilon\,R_{ab}{}^{ef}M_{ef}
+*E^cE^d\,\upsilon\Gamma_c\Gamma^{ab}\mathcal S\Gamma_d\Gamma_{11}\upsilon\,R_{ab}{}^{ef}M_{ef}\,.
\end{align}
Note that this term could be written as the $\mathcal J'\mathcal J'$-term in the supersymmetric case in eq. (\ref{eq:Jid2}). In that case it was canceled by a term in the curvature coming from $(Q\mathcal J')^2$. This is not possible in the non-supersymmetric case and since this term is generically non-zero the Lax connection constructed here does not work in the non-supersymmetric case. Integrability of the string in a non-supersymmetric symmetric space RR background therefore remains an open question.

\subsection{Supercoset Lax connections: $H_{abc}\neq0$ cases}
For the cases with non-zero NSNS flux we have not been able to find a general form of the Lax connection like we found for the case of zero NSNS flux. We will therefore construct the Lax connection case by case (actually it turns out that we can cover several cases with one construction) for the known supersymmetric symmetric space backgrounds with non-zero NSNS flux in table \ref{table:2}. We will describe only the Lax connection for the corresponding supercoset models here as it is clear that it can be extended to the full string by including the additional $\upsilon$ fermions just as we did in the case of zero NSNS flux since a Lax connection up to quadratic order in all fermions (i.e. $\vartheta$ and $\upsilon$) was already constructed in \cite{Wulff:2014kja}.

A supercoset Lax connection for $AdS_3\times S^3\times S^3\times S^1$ with mixed RR and NSNS flux was already constructed in \cite{Cagnazzo:2012se}. Since our construction, described below, should simply result in a Lax connection related to theirs by a gauge-transformation we will skip this case and concentrate on the cases for which a supercoset Lax connection has not previously been constructed.

\subsubsection{$AdS_2\times S^2\times S^2\times T^4$}
We will first consider $AdS_2\times S^2\times S^2\times T^4$ with a mix of RR and NSNS flux (related by S-duality to the pure RR flux case). For the details of the solution see \cite{Wulff:2014kja}. A natural guess for the Lax connection is the straightforward generalization of the bosonic Lax connection in eq. (\ref{eq:Lax}) namely (we have dropped the $E^{a'}P_{a'}$ term since it has no effect)
\begin{align}
L=&
\frac12\Omega^{\hat a\hat b}M_{\hat a\hat b}
+(1+\hat\alpha)E^{\hat a}P_{\hat a}+\hat\beta*E^{\hat a}P_{\hat a}
+\beta(*E^{a'}+\Lambda^{a'})P_{a'}
\nonumber\\
&{}
+\frac{\beta}{4}E^{a'}H_{a'\hat b\hat c}M^{\hat b\hat c}
+\frac{\alpha}{4}*E^{a'}H_{a'\hat b\hat c}M^{\hat b\hat c}
+QVE
\,,
\end{align}
where we have added a term involving the supersymmetry generators $Q$. Like in the previous section $V$ is a matrix depending on the spectral parameter whose form will be determined below. Recall that the index $a'$ runs over the flat $T^4$-directions while $\hat a$, in this case, runs over the rest and that the one-form $\Lambda^{a'}$ is defined so that $d\Lambda^{a'}=\frac12E^BE^AH_{AB}{}^{a'}$ in analogy with the bosonic case (the only purpose of the $P_{a'}$-term in the Lax connection is to encode the equation of motion of the flat directions).\footnote{As an explicit realization we can take 
$$
\Lambda^{a'}=h^{a'}{}_{\hat a\hat b}\Omega^{\hat a\hat b}\,,
$$
where $h_{a'bc}$ is defined so that $R_{bc}{}^{de}h_{a'de}=H_{a'bc}$. Indeed, using the fact that $G_{a'}E=0$ and eq. (\ref{eq:extra-ids}) we get
$$
d\Lambda^{a'}=\frac12E^{\hat b}E^{\hat a}H^{a'}{}_{\hat a\hat b}-\frac{i}{2}E\Gamma^{a'}\Gamma_{11}E=\frac12E^BE^AH_{AB}{}^{a'}\,.
$$
Let us also note for completeness that the $B$-field of this supercoset model can be constructed as
$$
B=E^{a'}\Lambda_{a'}-\frac{i}{4}E\mathcal K\Gamma_{11}E\,,
%
$$
as can be shown using similar identities.} The NSNS flux is of the form $H\sim\omega_2dy$ and satisfies \cite{Wulff:2014kja}
\begin{equation}
H_{\hat a\hat bc'}H^{c'\hat d\hat e}=-q^2R_{\hat a\hat b}{}^{\hat d\hat e}\,.
\end{equation}
Comparing to eqs. (\ref{eq:C2-H-cond}) and (\ref{eq:alphahat}) we have\footnote{To have a natural action of the $\mathbbm Z_4$-automorphism we should take instead the choice described in footnote \ref{foot:alt-hatted}.}
\begin{equation}
\hat\beta=\eta\beta\,,\quad1+\hat\alpha=\eta(2+\alpha)\,,\qquad\eta=\frac12\sqrt{\frac{2+q^2\alpha}{2+\alpha}}\,.
\end{equation}

When we compute the curvature of this Lax connection it is clear that the terms involving two bosonic supervielbeins will be of the right form since for these terms the calculation is exactly the same as in the bosonic case. This means that we only need to check the terms involving fermionic supervielbeins. Using the supercoset constraints on the torsion and curvature (see for example \cite{Wulff:2013kga})
\begin{equation}
T^a=-\frac{i}{2}E\Gamma^aE\,,\quad
R^{ab}=\frac{i}{8}EG^{ab}E+\frac12E^dE^cR_{cd}{}^{ab}\,,
%
\end{equation}
and the superisometry algebra in eq. (\ref{eq:Q-algebra2}) we find the following terms in the curvature of the Lax connection containing two fermionic supervielbeins (note that we must take into account the $E\Gamma^a\Gamma_{11}E$ term coming from the bosonic equations of motion)
\begin{align}
(dL-LL)_{0,2}=&
-\frac{i}{2}(1+\hat\alpha)E\Gamma^{\hat a}E\,P_{\hat a}
+\frac{i\hat\beta}{2}E\Gamma^{\hat a}\Gamma_{11}E\,P_{\hat a}
+\frac{i}{2}EV^\dagger\Gamma^{\hat a}VE\,P_{\hat a}
\nonumber\\
&{}
-\frac{i}{8}E\Gamma^{\hat a}\hat{\mathcal S}\Gamma^{\hat b}E\,M_{\hat a\hat b}
+\frac{i}{8}EV^\dagger\Gamma^{\hat a}\hat{\mathcal S}\Gamma^{\hat b}VE\,M_{\hat a\hat b}
+\frac{i\alpha}{8}E\Gamma_{a'}\Gamma_{11}E\,H^{a'\hat b\hat c}M_{\hat b\hat c}\,.
\label{eq:curv02-1}
\end{align}
Here we have used eqs. (\ref{eq:Gab-K}) and (\ref{eq:aprime-rel}) and the fact that $c_{[\hat a]}=2$. To prove the flatness of the Lax connection we will need, in addition to the general conditions derived so far, the following two conditions involving $\Gamma_{9'}=\hat q\Gamma_9-q\Gamma_{11}$ (recall that the $\mathbbm Z_4$-automorphism involves $\Gamma_{11'}=q\Gamma_9+\hat q\Gamma_{11}$ satisfying (\ref{eq:gamma11-rel}) and that $q^2+\hat q^2=1$)
\begin{equation}
q(\hat{\mathcal S}\Gamma_{a'}\Gamma_{9'})^{\hat\alpha}{}_{\hat\beta}=-H_{a'\hat b\hat c}(\Gamma^{\hat b\hat c})^{\hat\alpha}{}_{\hat\beta}\,,\qquad
(\hat{\mathcal S}\Gamma_{\hat a}\Gamma_{9'})^{\hat\alpha}{}_{\hat\beta}=0\,.
\label{eq:extra-ids}
\end{equation}
Multiplying the first first equation with $\mathcal K$ and contracting with $H^{a'\hat b\hat c}$ we find
\begin{equation}
8(C\Gamma_{a'}\Gamma_{9'})_{\hat\alpha\hat\beta}H^{a'\hat b\hat c}
=q(\mathcal K\Gamma^{\hat d\hat e})_{\hat\alpha\hat\beta}R_{\hat d\hat e}{}^{\hat b\hat c}
=4q(C\Gamma^{\hat b}\hat{\mathcal S}\Gamma^{\hat c})_{\hat\alpha\hat\beta}
\,,\qquad
\label{eq:EEH-EER-rel}
\end{equation}
which lets us write the last term in eq. (\ref{eq:curv02-1}) in a similar form as the previous two terms. Furthermore, since the matrix $V$ must satisfy $\mathcal PV\mathcal P=V$ there is essentially only one choice possible namely
\begin{equation}
V=a+b\Gamma_{11'}\,,\qquad V^\dagger=a-b\Gamma_{11'}\,,
\label{eq:V-ads2}
\end{equation}
for some constants $a$ and $b$ depending on the spectral parameter to be determined. This is the analog of the pure RR case, eq. (\ref{eq:V-RR}). Recall that $\Sigma=i\Gamma_{11'}$ was the matrix appearing in the $\mathbbm Z_4$-automorphism of the algebra discussed in sec. \ref{sec:semisymmetric}. Using the fact that $\Gamma_{11}=\hat q\Gamma_{11'}-q\Gamma_{9'}$ and the relations in eqs. (\ref{eq:gamma11-rel}) and (\ref{eq:extra-ids}) we finally find
\begin{align}
(dL-LL)_{0,2}=&
\frac{i}{2}(a^2+b^2-1-\hat\alpha)E\Gamma^{\hat a}E\,P_{\hat a}
+\frac{i}{2}(2ab+\hat q\hat\beta)E\Gamma^{\hat a}\Gamma_{11'}E\,P_{\hat a}
\nonumber\\
&{}
+\frac{i}{4}(a^2-b^2-1-\frac{q^2\alpha}{2})E\Gamma^{\hat a}\hat{\mathcal S}\Gamma^{\hat b}E\,M_{\hat a\hat b}\,.
\end{align}
These three terms have to vanish separately giving us three conditions on two unknowns. Luckily this overdetermined system has a solution namely
\begin{equation}
a=\frac12\Big(\sqrt{1+\hat\alpha-\hat q\hat\beta}+\sqrt{1+\hat\alpha+\hat q\hat\beta}\Big)\,,
\qquad b=\frac12\Big(\sqrt{1+\hat\alpha-\hat q\hat\beta}-\sqrt{1+\hat\alpha+\hat q\hat\beta}\Big)\,.
\label{eq:a-b}
\end{equation}
This means that $V$ satisfies similar identities to the zero NSNS flux case (eq. (\ref{eq:Vids}))
\begin{equation}
VV^\dagger=1+\frac{q^2\alpha}{2}\,,\qquad V^2=1+\hat\alpha-\hat q\hat\beta\Gamma_{11'}\,.
\label{eq:V-ids}
\end{equation}

It remains to check the terms in the curvature proportional to one bosonic and one fermionic supervielbein. Using the constraint on the fermionic torsion
\begin{equation}
T^\alpha=\frac18E^a\,(G_aE)^\alpha=\frac14E^{\hat a}\,(\hat{\mathcal S}\Gamma_{\hat a}E)^\alpha\,,
\end{equation}
these become
\begin{align}
(dL-LL)_{1,1}=&
\frac14E^{\hat a}\,QV\hat{\mathcal S}\Gamma_{\hat a}E
-\frac14((1+\hat\alpha)E^{\hat a}+\hat\beta*E^{\hat a})Q\hat{\mathcal S}\Gamma_{\hat a}VE
\nonumber\\
&{}
+\frac18(\beta E^{a'}+\alpha*E^{a'})H_{a'\hat b\hat c}\,Q\Gamma^{\hat b\hat c}VE\,.
\end{align}
Using eqs. (\ref{eq:aprime-rel}) and (\ref{eq:extra-ids}) together with the identities satisfied by $V$, eq. (\ref{eq:V-ids}), this can be written as
\begin{align}
(dL-LL)_{1,1}=&
-\frac{\hat\beta}{4}QV^\dagger\hat{\mathcal S}\Big(*E^a\Gamma_aE-E^a\Gamma_a\Gamma_{11}E\Big)
+\frac{q\alpha}{8}QV\hat{\mathcal S}\Gamma_{9'}\Big(*E^a\Gamma_aE-E^a\Gamma_a\Gamma_{11}E\Big)
\nonumber\\
&{}
+\frac{q}{8}E^{a'}\,Q[(\beta+\hat q\alpha\Gamma_{11'})V-2\hat\beta V^\dagger]\hat{\mathcal S}\Gamma_{9'}\Gamma_{a'}E\,.
\end{align}
Using the identities satisfied by $V$, we find that
\begin{equation}
(\beta+\hat q\alpha\Gamma_{11'})V-2\hat\beta V^\dagger=0
\end{equation}
so that the last term vanishes. The first term is proportional to the fermionic equation of motion of the supercoset model. At first glance then second term does not look like it is coming from the supercoset action since $\Gamma_{9'}$ does not commute with the projector $\mathcal P_1=\frac12(1+\Gamma^{6789'})$, where $\mathcal P=\mathcal P_1\mathcal P_2$ is the supersymmetry projector (see \cite{Wulff:2014kja}). However, recalling the fact that kappa symmetry implies that
\begin{equation}
\Gamma\Psi=\Psi\,,
\end{equation}
where $\Psi=0$ is the full fermionic equation of motion of the Green-Schwarz string and $\frac12(1+\Gamma)$ is the kappa symmetry projector, we have
\begin{equation}
0=\mathcal P_1\Psi=\mathcal P_1\Gamma\Psi
=\Gamma\Gamma_{11}\mathcal P_1\Gamma_{11}\Psi\,,
\end{equation}
where the last step holds provided that $\Gamma\Gamma_{11}\propto\varepsilon^{ij}E_i{}^aE_j{}^b\Gamma_{ab}$ commutes with $\mathcal P_1$. From the form of $\mathcal P_1$ we see that this is true as long as there is no motion in the $T^4$-directions. We conclude that at least for such configurations the equations following from flatness of the proposed Lax connection are indeed equivalent to the supercoset equations of motion.

This concludes the proof of the classical integrability of the supercoset model corresponding to $AdS_2\times S^2\times S^2\times T^4$ with mixed RR and NSNS flux. Note that this background corresponds to the supercoset
\begin{equation}
\frac{D(2,1;\alpha)}{SO(1,1)\times SO(2)\times SO(2)}\times U(1)^4\,,
\end{equation}
where the last $U(1)$ torus direction ($x^9$) decouples and can consistently be removed. The three remaining torus directions cannot be removed as $H$ has a leg along them and so they couple to the other coordinates through the equations of motion. Let us now consider the remaining backgrounds in table \ref{table:2}.

\subsubsection{$AdS_{2,3}\times S^{2,3}\times S^{2,3}\times T^{2,3}$}
These backgrounds have non-tunable NSNS flux on the two-dimensional factor(s) times a torus direction and are related by Hopf T-dualities to $AdS_3\times S^3\times S^3\times S^1$ with pure RR flux. The details of the solutions were given in \cite{Wulff:2014kja}. The construction of the supercoset Lax connection is very similar to the $AdS_2\times S^2\times S^2\times T^4$ case. We take
\begin{align}
L=&
\frac12\Omega^{\hat a\hat b}M_{\hat a\hat b}
+\Omega^{\tilde a\tilde b}\varepsilon_{\tilde a\tilde b\tilde c}P^{L\tilde c}
+(1+\hat\alpha)E^{\hat a}P_{\hat a}+\hat\beta*E^{\hat a}P_{\hat a}
+(1+\alpha)E^{\tilde a}P^L_{\tilde a}+\beta*E^{\tilde a}P^L_{\tilde a}
\nonumber\\
&{}
+\beta(*E^{a'}+\Lambda^{a'})P_{a'}
+\frac{\beta}{4}E^{a'}H_{a'\hat b\hat c}M^{\hat b\hat c}
+\frac{\alpha}{4}*E^{a'}H_{a'\hat b\hat c}M^{\hat b\hat c}
+QVE
\,,
\label{eq:Lax-T23}
\end{align}
where now $\hat a$ runs over the two-dimensional curved factors in the geometry and $\tilde a$ runs over the three-dimensional curved factors. The NSNS flux now satisfies
\begin{equation}
H_{\hat a\hat bc'}H^{c'\hat d\hat e}=-R_{\hat a\hat b}{}^{\hat d\hat e}\,,
\end{equation}
so that
\begin{equation}
1+\hat\alpha=1+\frac{\alpha}{2}\,,\qquad\hat\beta=\frac{\beta}{2}\,.
\end{equation}
These expressions are just the $AdS_2\times S^2\times S^2\times T^4$ ones with $q=1$. We will see that this holds for the other equations as well.

A difference compared to the bosonic Lax connection in eq. (\ref{eq:Lax}) is that the generators $P_{\tilde a},\,M_{\tilde a\tilde b}$ now only appear in the combination 
\begin{align}
P^L_{\tilde a}=\frac12(P_{\tilde a}+\frac14\varepsilon_{\tilde a\tilde b\tilde c}M^{\tilde b\tilde c})\,.
\end{align}
These generate the ''left'' $\mathfrak{su}(2)$ ($\mathfrak{sl}(2)$) of the $SO(4)$ ($SO(2,2)$) isometries of $S^3$ ($AdS_3$) and satisfy
\begin{equation}
[P^L_{\tilde a},P^L_{\tilde b}]=\frac12\varepsilon_{\tilde a\tilde b\tilde c}P^{L\tilde c}\,.
\end{equation}
Note that the ''right'' $\mathfrak{su}(2)$ ($\mathfrak{sl}(2)$) decouples, in the sense that
\begin{equation}
[P^R_{\tilde a},P^L_{\tilde b}]=0\,,\qquad [P^R_{\tilde a},Q]=0\,,
\end{equation}
where $P^R$ is defined like $P^L$ but with the opposite sign for the $M_{\tilde a\tilde b}$ term. Keeping only the $P^L$-term in the bosonic Lax connection in eq. (\ref{eq:Lax}) does not change the analysis there since the $P^R$ terms decouple. Therefore the terms in the curvature proportional to two bosonic supervielbeins will again vanish modulo the equations of motion.

Looking at the terms with two fermionic supervielbeins we find
\begin{align}
(dL-LL)_{0,2}=&
-\frac{i}{2}(1+\frac{\alpha}{2})E\Gamma^{\hat a}E\,P_{\hat a}
+\frac{i\beta}{4}E\Gamma^{\hat a}\Gamma_{11}E\,P_{\hat a}
+\frac{i}{2}EV^\dagger\Gamma^{\hat a}VE\,P_{\hat a}
\nonumber\\
&{}
-\frac{i}{2}(1+\alpha)E\Gamma^{\tilde a}E\,P^L_{\tilde a}
+\frac{i\beta}{2}E\Gamma^{\tilde a}\Gamma_{11}E\,P^L_{\tilde a}
+\frac{i}{2}EV^\dagger\Gamma^{\tilde a}VE\,P_{\tilde a}
\nonumber\\
&{}
-\frac{i}{8}E\Gamma^{\hat a}\hat{\mathcal S}\Gamma^{\hat b}E\,M_{\hat a\hat b}
+\frac{i}{8}EV^\dagger\Gamma^{\hat a}\hat{\mathcal S}\Gamma^{\hat b}VE\,M_{\hat a\hat b}
+\frac{i\alpha}{8}E\Gamma_{a'}\Gamma_{11}E\,H^{a'\hat b\hat c}M_{\hat b\hat c}
\nonumber\\
&{}
-\frac{i}{8}E\Gamma^{\tilde a}\hat{\mathcal S}\Gamma^{\tilde b}E\,\varepsilon_{\tilde a\tilde b\tilde e}P^{L\tilde e}
+\frac{i}{16}EV^\dagger\Gamma^{\tilde a}\hat{\mathcal S}\Gamma^{\tilde b}VE\,M_{\tilde a\tilde b}\,.
\end{align}
where we used the fact that $c_{[\hat a]}=2$ and $c_{[\tilde a]}=1$. Taking $V$ to be given by the expression in eqs. (\ref{eq:V-ads2}) and (\ref{eq:a-b}) but with $q=1$ we find
\begin{equation}
V=V^\dagger=\sqrt{1+\frac{\alpha}{2}}\,.
\end{equation}
Using this and the fact that the identities in eq. (\ref{eq:extra-ids}) still hold (the last one holds also with $\hat a\rightarrow\tilde a$), but with $q=1$ so that $\Gamma_{9'}=-\Gamma_{11}$ and $\Gamma_{11'}=\Gamma_9$, the terms involving hatted indices cancel in the same way as in the previous section and we are left with
\begin{align}
(dL-LL)_{0,2}=&
-\frac{i}{2}(1+\alpha)E\Gamma^{\tilde a}E\,P^L_{\tilde a}
-\frac{i}{8}E\Gamma^{\tilde a}\hat{\mathcal S}\Gamma^{\tilde b}E\,\varepsilon_{\tilde a\tilde b\tilde e}P^{L\tilde e}
\nonumber\\
&{}
+\frac{i}{4}(2+\alpha)E\Gamma^{\tilde a}E\,P_{\tilde a}
+\frac{i}{32}(2+\alpha)E\Gamma^{\tilde a}\hat{\mathcal S}\Gamma^{\tilde b}E\,M_{\tilde a\tilde b}\,.
\end{align}
Using the additional relation
\begin{equation}
E\Gamma^{[\tilde a}\hat{\mathcal S}\Gamma^{\tilde b]}E
=
2E\Gamma_{\tilde c}E\,\varepsilon^{\tilde a\tilde b\tilde c}
\label{eq:extra-ids2}
%
\end{equation}
and the form of $P^L_{\tilde a}$ we see that the remaining terms indeed cancel.

It remains to analyze the terms with one bosonic and one fermionic supervielbein or, equivalently, the terms in the curvature of $L$ involving $Q$. Using
\begin{equation}
T^\alpha=\frac18E^a\,(G_aE)^\alpha=\frac14E^{\hat a}\,(\hat{\mathcal S}\Gamma_{\hat a}E)^\alpha+\frac18E^{\tilde a}\,(\hat{\mathcal S}\Gamma_{\tilde a}E)^\alpha
\end{equation}
together with the relations in eqs. (\ref{eq:aprime-rel}) and (\ref{eq:extra-ids}) these take the form
\begin{align}
(dL-LL)_{1,1}=&
-\frac{\beta}{8}QV\hat{\mathcal S}\Big(*E^{\hat a}\Gamma_{\hat a}+*E^{\tilde a}\Gamma_{\tilde a}E-E^{a'}\Gamma_{a'}\Gamma_{11}E\Big)
\nonumber\\
&{}
-\frac{\alpha}{8}QV\hat{\mathcal S}\Gamma_{11}\Big(*E^{a'}\Gamma_{a'}E-E^{\hat a}\Gamma_{\hat a}\Gamma_{11}E-E^{\tilde a}\Gamma_{\tilde a}\Gamma_{11}E\Big)\,.
\end{align}
The first term is proportional to the fermionic equation of motion of the supercoset model. The second term looks different since $\Gamma_{11}$ does not commute with $\mathcal P_1=\frac12(1+\Gamma^9)$ ($T^2$-cases) or $\mathcal P_1=\frac12(1+\Gamma^{0123456})$ ($T^3$-cases) appearing in the supersymmetry projector $\mathcal P$ (see \cite{Wulff:2014kja}). But, just as we saw in the last section we have
\begin{equation}
0=\mathcal P_1\Psi=\mathcal P_1\Gamma\Psi=\Gamma\Gamma_{11}\mathcal P_1\Gamma_{11}\Psi\,,
\end{equation}
where $\frac12(1+\Gamma)$ is the kappa symmetry projector and $\Psi=0$ is the fermionic equation of motion, provided that $\Gamma\Gamma_{11}\propto\varepsilon^{ij}E_i{}^aE_j{}^b\Gamma_{ab}$ commutes with $\mathcal P_1$, i.e. as long as there is no motion in the $x^9$-direction for the $T^2$-cases or in the $T^3$-directions. Therefore we conclude that at least for such configurations of the string the flatness of the proposed Lax connection is indeed equivalent to the equations of motion.

This completes the proof of the classical integrability of the corresponding supercoset models, e.g.
\begin{align}
AdS_2\times S^2\times S^3\times T^3:&\qquad\frac{D(2,1;\alpha)\times SU(2)}{SO(1,1)\times SO(2)\times SO(3)}\times U(1)^3\nonumber\\
AdS_3\times S^3\times S^2\times T^2:&\qquad\frac{D(2,1;\alpha)\times SL(2,\mathbbm R)\times SU(2)}{SO(2,1)\times SO(3)\times SO(2)}\times U(1)^2\nonumber\,.
\end{align}

Let us note that since all identities we used in this section were the same as in the last section but with $q=1$ and the Lax connections had the same form (modulo the fact that the $\tilde a$-directions were absent in the first case) we can in fact describe all the $AdS_{2,3}\times S^{2,3}\times S^{2,3}\times T^{2,3,4}$ cases in table \ref{table:2} with the Lax connection in eq. (\ref{eq:Lax-T23}).

\section{Conclusions}
We have seen how the well known integrability of the bosonic string on a symmetric space, described by a symmetric space sigma model, extends also to the case of non-zero NSNS flux which respects the isometries provided the latter "squares" to the Riemann tensor. This condition cannot be the most general condition for integrability however, as we have remarked, and an interesting open problem is to find the most general condition on $H$ needed for integrability. Our main concern here was however the integrability of the superstring on symmetric spaces. In the case of zero NSNS flux we have shown the integrability of the type II superstring by showing that it can be viewed as an extension of a supercoset sigma model by additional fermions and showing that this extension preserves the integrability of the original supercoset model. We demonstrated this explicitly up to quadratic order in the non-coset fermions and we believe it to hold to all orders though a proof is so far lacking. We proved the integrability only in the case where there is some non-zero amount of supersymmetry preserved. The non-supersymmetric case deserves further study.

In the case of non-zero NSNS flux the integrability question is more complicated and we have not been able to find a general construction of the Lax connection for the superstring. Therefore we had to settle for showing that a similar Lax connection construction as for the zero NSNS flux case exist for the known supersymmetric symmetric space backgrounds with NSNS flux (table \ref{table:2}). Here we constructed explicitly the corresponding supercoset Lax connections, which have not appeared in the literature before, and noted that it must be possible to extend them by the non-coset fermions since an alternative Lax connection was found to second order in all the fermions in \cite{Wulff:2014kja}.

Some results derived on the way include a proof that a (type II) symmetric space supergravity background which preserves some supersymmetry has a supercoset sub-superspace. We also showed that in the absence of NSNS flux this is in fact a semisymmetric superspace. This is also true for some NSNS flux backgrounds (the ones related by S-duality to backgrounds without NSNS flux) but not for others. This can therefore not be a necessary condition for (classical) integrability. We have also derived the form of the superisometry algebra in terms of the (constant) fluxes for a (type II) symmetric space supergravity background. These results should prove useful in further studies of strings on symmetric spaces and in particular of their integrability properties.

\section*{Acknowledgments}
I wish to thank A. Tseytlin for useful discussions. This work was supported by the ERC Advanced grant No.290456 ``Gauge theory -- string theory duality''.

\

\appendix

\section{Expansion of supergeometry and superisometries to order $\upsilon^2$}\label{app:J}
In \cite{Wulff:2013kga} the supergeometry of a general type II supergravity background was determined up to fourth order in the fermions $\Theta$ by solving the differential equations for the $\Theta$-dependence order by order starting from the bosonic geometry at order $\Theta^0$. In the special case of interest here where we have an underlying supercoset geometry we can instead start from this geometry and compute the expansion in the non-coset fermions $\upsilon$. The calculation is essentially identical to that described in \cite{Wulff:2013kga} with $\Theta\rightarrow\upsilon$ except now the fermionic supervielbein is non-zero in the lowest order and we refer to that paper for the details. The $\upsilon$-expansion of the supervielbeins and spin connection become, to the order we will need here,
\begin{equation}
\u E^a=E^a-iE\Gamma^a\upsilon-\frac{i}{2}\mathcal D\upsilon\Gamma^a\upsilon+\mathcal O(\upsilon^3)\,,\quad
\u E^\alpha=E^\alpha+(\mathcal D\upsilon)^\alpha+\mathcal O(\upsilon^2)\,,\quad
\u\Omega^{ab}=\Omega^{ab}-\frac{i}{4}\upsilon G^{ab}E+\mathcal O(\upsilon^2)\,,
\end{equation}
where
\begin{equation}
\mathcal D\upsilon=\big(d-\frac14\Omega^{ab}\Gamma_{ab}+\frac18E^a\,G_a\big)\upsilon\,.
\end{equation}

Since the superisometry Noether current defined in eq. (\ref{eq:calJ}) involves also the Killing vector and Killing spinor superfields we will also need their expansion. In fact this is determined in a very similar way to the expansion of the supervielbeins and spin connection. The expansion in $\Theta$, starting from the bosonic expressions, was found in \cite{Wulff:2014kja} and adapting it to the expansion in $\upsilon$ starting from the supercoset expressions essentially amounts to replacing $\Theta\rightarrow\upsilon$ and we find, to the order we will need here,
\begin{align}
&\u K_a=K_a+i\upsilon\Gamma_a\xi+\frac{i}{16}\upsilon\Gamma_aG_b\upsilon\,K^b-\frac{i}{8}\upsilon\Gamma_{abc}\upsilon\,\nabla^bK^c+\mathcal O(\upsilon^3)\,,
\nonumber\\
&\u\xi^\alpha=\xi^\alpha+\frac18(G_a\upsilon)^\alpha\,K^a-\frac14(\Gamma^{ab}\upsilon)^\alpha\,\nabla_aK_b+\mathcal O(\upsilon^2)\,,\qquad
\u\nabla_a\u K_b=\nabla_aK_b-\frac{i}{4}\upsilon G_{ab}\xi+\mathcal O(\upsilon^2)\,.
\end{align}

The Noether current defined in eq. (\ref{eq:calJ}) also involves the object $\u\Lambda$ satisfying
\begin{equation}
d\u\Lambda=i_{\u K}\u H=\frac12\u E^b\u E^a\u H_{abc}\u K^c-i\u E^a\,\u E\Gamma_a\Gamma_{11}\u\xi-\frac{i}{2}\u E\Gamma_a\Gamma_{11}\u E\,\u K^a\,.
\end{equation}
Its expansion in $\upsilon$ is given by (this is analogous to how the expansion of $B$ was found from that of $H$ in \cite{Wulff:2013kga})
\begin{eqnarray}
\u\Lambda=
\Lambda+\int_0^1dt\,(i_\upsilon i_{\u K}\u H)_t
=
\Lambda-i\int_0^1dt\,\left(\u E^a\,\upsilon\Gamma_a\Gamma_{11}\u\xi-\upsilon\Gamma_a\Gamma_{11}\u E\,\u K^a\right)_t\,,
\end{eqnarray}
where the subscript $t$ means that $\upsilon$ is replaced by $t\upsilon$ in the expansion of the superfields. The lowest, supercoset, component satisfies
\begin{equation}
d\Lambda=i_KH=\frac12E^bE^aH_{abc}K^c-iE^a\,E\Gamma_a\Gamma_{11}\xi-\frac{i}{2}E\Gamma_a\Gamma_{11}E\,K^a\,.
\end{equation}
When $H_{abc}=0$ we can take\footnote{Note that this is just $-i_KB$ where the $B$-field of the supercoset model is given by
$$
B=-\frac{i}{4}E\mathcal K\Gamma_{11}E\,,
$$
as is easily verified.
}
\begin{equation}
\Lambda=\frac{i}{2}E\mathcal K\Gamma_{11}\xi\,.
\end{equation}
Indeed, using eq. (\ref{eq:nabla-xi}), the torsion constraint eq. (\ref{eq:supercoset-constr}) and the symmetry properties of the gamma matrices we find (recalling that in this case $\mathcal K\mathcal S=8\mathcal PC$ and $[\mathcal P,\Gamma_{11}]=0$)
\begin{equation}
d\Lambda=-iE^a\,E\Gamma_a\Gamma_{11}\xi-\frac{i}{2}E\Gamma_a\Gamma_{11}E\,K^a\,.
\end{equation}

Using the expansions of the supervielbeins, Killing vector and Killing spinor superfields in $\upsilon$ we can now find the expansion of the superisometry Noether current defined in eq. (\ref{eq:calJ}). Splitting it into components according to eq. (\ref{eq:Jsplit}) we find for $H_{abc}=0$
\begin{align}
\mathcal J^a=&
E^a
-iE\Gamma^a\upsilon
+i\upsilon\Gamma^a\Gamma_{11}*E
+\frac{i}{16}E^b\,\upsilon\Gamma_b\mathcal S\Gamma^a\upsilon
-\frac{i}{16}*E^b\,\upsilon\Gamma_b\Gamma_{11}\mathcal S\Gamma^a\upsilon
\nonumber\\
&{}
-\frac{i}{2}\mathcal D\upsilon\Gamma^a\upsilon
+\frac{i}{2}\upsilon\Gamma^a\Gamma_{11}*\mathcal D\upsilon
+\mathcal O(\upsilon^3)\,,
\\
\mathcal J^{ab}=&
\frac{i}{8}E^e\,\upsilon\Gamma_e{}^{cd}\upsilon\,R_{cd}{}^{ab}
-\frac{i}{8}*E^e\,\upsilon\Gamma_e{}^{cd}\Gamma_{11}\upsilon\,R_{cd}{}^{ab}
+\mathcal O(\upsilon^3)\,,
\\
\mathcal J^{\hat\alpha}=&
-\frac12(\Gamma_{11}*E)^{\hat\alpha}
-\frac18E^a\,(\mathcal S\Gamma_a\upsilon)^{\hat\alpha}
+\frac18*E^a\,(\mathcal S\Gamma_a\Gamma_{11}\upsilon)^{\hat\alpha}
+\frac{i}{8}E\Gamma^a\upsilon\,(\mathcal S\Gamma_a\upsilon)^{\hat\alpha}
\nonumber\\
&{}
-\frac{i}{16}*E\Gamma^a\upsilon\,(\mathcal S\Gamma_a\Gamma_{11}\upsilon)^{\hat\alpha}
-\frac{i}{16}\upsilon\Gamma^a\Gamma_{11}*E\,(\mathcal S\Gamma_a\upsilon)^{\hat\alpha}
+\mathcal O(\upsilon^3)\,.
\end{align}
Below we will prove some relations satisfied by $\mathcal J$ which are needed to prove the integrability. The ultimate reason for these relations should be the kappa symmetry of the string. It would be very interesting to derive them directly from the requirement of kappa symmetry but we will leave this for future work. Here we will instead just verify them by a direct calculation using the above expansions.

\subsection{Relations satisfied by components of $\mathcal J$}
In this section we set $H_{abc}=0$ since this is the only case we need here. We also assume that some supersymmetry is preserved. To derive the first relation we consider the following two terms
\begin{align}
&{}
8\upsilon\Gamma_e{}^{cd}\upsilon\,R_{cd}{}^{ab}(\Gamma_{ab}E)
+8\upsilon\Gamma_e{}^{cd}\Gamma_{11}\upsilon\,R_{cd}{}^{ab}(\Gamma_{ab}\Gamma_{11}E)
\nonumber\\
&=
\upsilon\Gamma_{ecd}\upsilon\,(\mathcal S\Gamma^c\mathcal S\Gamma^dE)
+\upsilon\Gamma_{ecd}\Gamma_{11}\upsilon\,(\mathcal S\Gamma^c\mathcal S\Gamma^d\Gamma_{11}E)
\nonumber\\
&=
-\upsilon\Gamma_{ec}\Gamma_{11}
\Big[
\Gamma_d\Gamma_{11}\upsilon\,(\mathcal S\Gamma^c\mathcal S\Gamma^dE)
-\Gamma_d\Gamma_{11}E\,(\mathcal S\Gamma^c\mathcal S\Gamma^d\upsilon)
-\Gamma_d\Gamma_{11}\mathcal S\Gamma^c\mathcal SC\,\upsilon\Gamma^dE
\nonumber\\
&{}\qquad\qquad
+\Gamma_d\upsilon\,(\mathcal S\Gamma^c\mathcal S\Gamma^d\Gamma_{11}E)
-\Gamma_dE\,(\mathcal S\Gamma^c\mathcal S\Gamma^d\Gamma_{11}\upsilon)
-\Gamma_d\mathcal S\Gamma^c\mathcal SC\,\upsilon\Gamma^d\Gamma_{11}E
\Big]
\nonumber\\
&{}\qquad
+\upsilon\Gamma_{ec}\Gamma_dE\,(\mathcal S\Gamma^c\mathcal S\Gamma^d\upsilon)
+\upsilon\Gamma^dE\,(\mathcal S\Gamma^c\mathcal S\Gamma_d\Gamma_{ec}\upsilon)
-\upsilon\Gamma_{ec}\Gamma_{11}\Gamma_dE\,(\mathcal S\Gamma^c\mathcal S\Gamma^d\Gamma_{11}\upsilon)
\nonumber\\
&{}\qquad
+\upsilon\Gamma^d\Gamma_{11}E\,(\mathcal S\Gamma^c\mathcal S\Gamma_d\Gamma_{ec}\Gamma_{11}\upsilon)\,.
\end{align}
The expression in square brackets vanishes by the basic Fierz identity eq. (\ref{eq:Fierz}). Using this and the expansion of $\mathcal J$ we find after a bit of algebra
\begin{align}
*\mathcal J^{ab}\,\Gamma_{ab}\Gamma_{11}E
-\mathcal J^{ab}\,\Gamma_{ab}E
=&
-\frac12E^a\,\mathcal S\Gamma_a\Gamma_{11}*\mathcal J'
+\frac12\mathcal J^a\,\mathcal S\Gamma_a\mathcal J'
\nonumber\\
&{}
-\frac{i}{64}\upsilon\Gamma^a\Gamma^b\Gamma_{11}[*E^c\,\Gamma_cE-E^c\,\Gamma_c\Gamma_{11}E]\,(\mathcal S\Gamma_a\mathcal S\Gamma_b\upsilon)
\nonumber\\
&{}
+\frac{i}{64}\upsilon\Gamma^a\Gamma^b[*E^c\,\Gamma_cE-E^c\,\Gamma_c\Gamma_{11}E]\,(\mathcal S\Gamma_a\mathcal S\Gamma_b\Gamma_{11}\upsilon)
+\mathcal O(\upsilon^3)\,,
\end{align}
where $\mathcal J'$ was defined in (\ref{eq:Jprime}) and we've used the fact that the dilatino equation in eq. (\ref{eq:inv-susy}) together with the fact that $\mathcal S=\mathcal P\mathcal S$ implies $\Gamma^a\mathcal S\Gamma_a\mathcal S=0=\mathcal S\Gamma^a\mathcal S\Gamma_a$ (recall that we set $H_{abc}=0$). The last two terms are proportional to the equations of motion (evaluated at $\upsilon=0$). This completes the derivation of the first condition.

To derive the second condition we compute, using the expansion of $\mathcal J$ in the previous section and the conditions from supersymmetry,
\begin{align}
\nabla\mathcal J^{ab}=&
-\frac{i}{8}\nabla*E^e\,\upsilon\Gamma_e{}^{cd}\Gamma_{11}\upsilon\,R_{cd}{}^{ab}
+\frac{1}{16}E\Gamma^eE\,\upsilon\Gamma_e{}^{cd}\upsilon\,R_{cd}{}^{ab}
+\frac{i}{4}E^e\,\upsilon\Gamma_e{}^{cd}\nabla\upsilon\,R_{cd}{}^{ab}
\nonumber\\
&{}
-\frac{i}{4}*E^e\,\upsilon\Gamma_e{}^{cd}\Gamma_{11}\nabla\upsilon\,R_{cd}{}^{ab}
\nonumber\\
=&
\frac{i}{4}\upsilon\Gamma^{cd}\Gamma_{11}
\Big(
*(E^e-iE\Gamma^e\upsilon)\,\Gamma_eE
+*E^e\,\Gamma_e\mathcal D\upsilon
-(E^e-iE\Gamma^e\upsilon)\,\Gamma_e\Gamma_{11}E
-E^e\,\Gamma_e\Gamma_{11}\mathcal D\upsilon
\Big)
\,R_{cd}{}^{ab}
\nonumber\\
&{}
-\frac{i}{8}(\nabla*E^e-\frac{i}{2}E\Gamma^e\Gamma_{11}E)\,\upsilon\Gamma_e{}^{cd}\Gamma_{11}\upsilon\,R_{cd}{}^{ab}
+\frac{i}{4}E\Gamma^{[a}\mathcal S\Gamma_{11}\Gamma^{b]}*\mathcal J'
+\frac{i}{8}\mathcal J'\Gamma^a\mathcal S\Gamma^b\mathcal J'
\nonumber\\
&{}
+\frac12\mathcal J^c\mathcal J^dR_{cd}{}^{ab}
-\frac12E^cE^dR_{cd}{}^{ab}
+\frac{1}{16}E\Gamma^eE\,\upsilon\Gamma_e{}^{cd}\upsilon\,R_{cd}{}^{ab}
+\frac18\upsilon\Gamma^eE\,E\Gamma_e\Gamma^{cd}\upsilon\,R_{cd}{}^{ab}
\nonumber\\
&{}
+\frac{1}{16}E\Gamma^e\Gamma_{11}E\,\upsilon\Gamma_e{}^{cd}\Gamma_{11}\upsilon\,R_{cd}{}^{ab}
+\frac18\upsilon\Gamma^e\Gamma_{11}E\,E\Gamma_e\Gamma^{cd}\Gamma_{11}\upsilon\,R_{cd}{}^{ab}
+\mathcal O(\upsilon^3)\,.
\end{align}
The last four terms cancel by the basic Fierz identity eq. (\ref{eq:Fierz}). The first term is proportional to the fermionic equation of motion and the second to the bosonic equation of motion. Again we have used the conditions from supersymmetry in eq. (\ref{eq:inv-susy}). This completes the derivation of the second condition.

To summarize we have shown that the components of the conserved current satisfy the following additional identities (on-shell)
\begin{align}
*\mathcal J^{ab}\,\Gamma_{ab}\Gamma_{11}E
-\mathcal J^{ab}\,\Gamma_{ab}E
+\frac12E^a\,\mathcal S\Gamma_a\Gamma_{11}*\mathcal J'
-\frac12\mathcal J^a\,\mathcal S\Gamma_a\mathcal J'
=&\,
\mathcal O(\upsilon^3)
\label{eq:Jid1}
\\
\nabla\mathcal J^{ab}
-\frac{i}{4}E\Gamma^{[a}\mathcal S\Gamma_{11}\Gamma^{b]}*\mathcal J'
-\frac{i}{8}\mathcal J'\Gamma^a\mathcal S\Gamma^b\mathcal J'
-\frac12\mathcal J^c\mathcal J^dR_{cd}{}^{ab}
+\frac12E^cE^dR_{cd}{}^{ab}
=&\,
\mathcal O(\upsilon^3)\,.
\label{eq:Jid2}
\end{align}

\end{document}